\documentclass[11pt]{article}

\usepackage[margin=1in]{geometry}
\usepackage{setspace}
\usepackage{amsmath}
\usepackage{amsfonts}
\usepackage{amsthm}
\usepackage{graphicx}
\usepackage{booktabs}
\usepackage{algorithm}
\usepackage{algorithmic}
\usepackage{url}
\usepackage{hyperref}
\usepackage{cleveref}
\usepackage{xcolor}
\usepackage{multirow}
\usepackage{subcaption}
\usepackage{enumitem}
\usepackage{natbib}

\hypersetup{
    colorlinks=true,
    linkcolor=blue,
    filecolor=magenta,
    urlcolor=cyan,
    citecolor=blue,
    pdftitle={Falsifying Discriminant Validity of Predictive Algorithms},
}

\theoremstyle{definition}
\newtheorem{definition}{Definition}[section]

\title{Falsifying Discriminant Validity of Predictive Algorithms}

\author{
Amanda Coston\\
Statistics Department\\
University of California, Berkeley\\
Berkeley, CA, USA\\
\texttt{acoston@berkeley.edu}
}

\date{}

\begin{document}

\maketitle

\begin{abstract}
Empirical investigations into unintended model behavior often show that the algorithm is predicting another outcome than what was intended.
These exposés highlight the need to identify when algorithms predict unintended quantities — ideally before deploying them into consequential settings.
We propose a falsification framework that provides a principled statistical test for \emph{discriminant validity}: the requirement that an algorithm predict intended outcomes better than impermissible ones.
Drawing on falsification practices from other fields, our framework compares calibrated prediction losses across outcomes to assess whether the algorithm exhibits discriminant validity with respect to a specified impermissible proxy.
In settings where the target outcome is difficult to observe, multiple permissible proxy outcomes may be available; our framework accommodates both this setting and the case with a single permissible proxy.
Throughout we use nonparametric hypothesis testing methods that make minimal assumptions on the data-generating process.
We illustrate the method in an admissions setting, where the framework establishes discriminant validity with respect to gender but fails to establish discriminant validity with respect to race.
This demonstrates how falsification can serve as an early validity check.
We also provide analysis in a criminal justice setting, where we highlight the limitations of our framework and emphasize the need for complementary approaches to assess other aspects of construct validity and external validity.
\end{abstract}


\section{Introduction}
\label{sec:introduction}

Predictive algorithms inform high-stakes decisions in domains ranging from healthcare to public services \citep{barocas2023fairness, eubanks2018automating, chouldechova2018case,obermeyer2019dissecting}.
In many of these settings, recent investigations raised concerns that algorithms operated in unintended ways, disproportionately impacting sensitive or protected groups such as people with disabilities and people identifying as Black \citep{mehrabi2021survey, obermeyer2019dissecting, vyas2020hidden, gilman2020welfare, charette2018unemployment}.
These cases have called into question the adequacy of standard evaluation practices, illustrating the need for rigorous pre-deployment evaluations that anticipate problems.\looseness=-1

Often the intended target is not observable, so evaluations instead use readily available proxy outcomes.
In hiring, algorithms are intended to assess whether an applicant would be successful if hired.
Since success is subjective and hard to measure, evaluations often use prior hiring decisions as a proxy outcome.
Unfortunately, proxy outcomes can encode historical biases.
The notorious Amazon recruiting algorithm penalized applications from women's colleges \citep{dastin2018amazon}.
Similarly, an algorithm used to aid admissions decisions at St.
George’s Medical School was revoked because, having been trained on prior admissions decisions, it learned to screen out women and racial minorities \citep{barocas2016big}.
In social services, algorithms used for fraud detection are intended to flag fraudulent welfare claims, but fraud can be hard to ascertain because it requires knowledge of the intent.
The UK Department of Work and Pensions used as a proxy outcome whether the claim included an inaccuracy, which included both intentional fraud and unintentional errors.
Independent investigations later uncovered that the algorithm made systematically more errors along demographic attributes including disability and age \citep{booth2024guardian,booth2024dwp}.
In child welfare, screening algorithms have come under scrutiny for alleged bias, raising the question whether algorithms are simply flagging families who interact with social services more \citep{ho2022oregon,ho2023pittsburgh}. \looseness=-1

Common to all these cases is the question \emph{Does the algorithm predict a different quantity than what we intend?} This concerns \emph{construct validity} \citep{cronbach1955construct,coston2023validity}---the degree to which the algorithm predicts the construct that was intended.
Construct validity involves several key components, of which the most relevant to our setting is discriminant validity \citep{campbell1959convergent}.
Discriminant validity requires that we establish that the measure does not capture constructs it is not intended to measure.
When proxy outcomes reflect social processes---such as prior hiring decisions---the algorithm may inadvertently learn to predict unintended and even impermissible quantities, but standard evaluations against that same proxy can fail to detect this fundamental flaw.
In such settings standard evaluations risk validating algorithms that are accurate with respect to the proxy but invalid with respect to the intended target.

Standard evaluation approaches compute predictive performance measures like accuracy or ROC curves, telling us how well the algorithm predicts the outcome used in evaluation. When this outcome is just a proxy for the desired outcome, these performance metrics may provide a misleading picture of how well the algorithm predicts the quantity we actually care about. Beyond standard performance metrics, there are evaluation approaches that leverage measurement error modeling and label bias frameworks \citep{guerdan2023ground,fawkes2024fragility,mhasawade2024causal}. Literature in this area provides methods for analyzing how performance metrics and fairness properties might vary under structural assumptions on the relationship between the proxy and intended outcome. None of the available methods, to our knowledge, assess whether the algorithm predicts a different quantity than what we intend.

In this paper, we propose a novel evaluation framework designed to assess whether the algorithm predicts a different quantity than what we intend.
We refer to such unintended quantities as \emph{impermissible proxies}---variables that are ethically or functionally inappropriate as prediction targets, such as race or gender.
Drawing on the established practice of falsification in fields such as statistics, psychometrics, and econometrics, our framework provides a principled statistical test for discriminant validity: whether the algorithm predicts permissible outcomes better than impermissible proxies.
The method requires only observable proxy outcomes and does not assume access to the unobservable intended target, making it broadly applicable across domains where the true quantity of interest is difficult to define or measure.

Throughout the paper we use two running examples to illustrate our falsification framework. Our first example, already introduced above, concerns predictive algorithms used to inform admissions decisions. The admissions task is to predict whether an applicant would thrive in the program if admitted. 
Success in the program is subjective, multifaceted and therefore hard to measure.
Possible proxies for success include GPA or employment upon graduation. To avoid the mistakes of previous admissions algorithms, we want to establish that the algorithm is not predicting an impermissible quantity such as gender or race \citep{barocas2016big}. Our second example is medical diagnosis. Suppose a researcher is interested in developing a new tool to diagnose depression, but they want to establish that their tool is not merely identifying people with comorbidities like anxiety or chronic pain \citep{li2024depression,hewitt1991beck}. This is important to establish for proper clinical care because the optimal treatment may be differ for chronic pain versus depression. We are then interested in how to assess whether the algorithm is (not) predicting an impermissible quantity such as a comorbidity.

Our falsification framework assesses one specific aspect of the validity of probabilistic binary classifiers—discriminant validity.
We emphasize, however, that validity is multifaceted, and our method does not address other important dimensions, including convergent validity, external validity, and content validity, among others \citep{cronbach1955construct,coston2023validity}.
We recommend using our method in conjunction with complementary methods for assessing other aspects of validity. 

Our falsification approach is not a fairness metric or test, although it has implications for fairness in certain settings.
Importantly, because our approach is validity-centered rather than metric-centered, it avoids several challenges that arise in standard fairness analyses.
In particular, we need not debate which performance metric to require parity on or how much discrepancy in that metric is allowable \citep{chouldechova2017fair,kleinberg2017inherent}.
Moreover, centering validity — and not a particular fairness metric — fundamentally changes the implication of the analysis.
When a disparity is observed in a fairness metric, a common but controversial remediation involves learning a model to minimize a constrained loss function that penalizes fairness violations \citep{lipton2018does,narayanan2026limits,barocas2023fairness,chouldechova2018frontiers}.
By contrast, failure to establish validity calls into question the foundations of the predictive task, implying that algorithm developers must return to the problem formulation phase to scrutinize design choices, the limitations of the data, and even the suitability of this task for prediction \citep{coston2023validity, passi2019problem}.\looseness=-1 

\textbf{Contributions:} (1) We synthesize the practice of falsification checks in other domains including statistics, social sciences, and econometrics to inform the development of our falsification framework (Section~\ref{sec:falsification in other domains}). (2)  We formulate the problem of falsifying the discriminant validity of predictive algorithms for probabilistic binary classifiers (Section~\ref{sec:problem_formulation}). (3) We propose a general framework for falsifying predictive algorithms and show how our framework can leverage multiple proxy outcomes. (4) We illustrate the real-world applicability of the method in an admissions setting and a criminal justice setting.

\section{Related Work}
\label{sec:related_work}


\paragraph{Validity in Algorithmic Systems.}
Our work builds on a growing literature that applies validity concepts from social science to algorithmic systems.
The foundational concepts of construct validity and discriminant validity originate in psychometrics \citep{cronbach1955construct,campbell1959convergent}.
Recent work argues that computational systems often assess unobservable constructs via observable proxies, and that mismatches in operationalization---a source of many fairness-related harms---mean models may not assess what we intend \citep{jacobs2021measurement,coston2023validity,salaudeen2025measurement,freiesleben2025benchmarking,wang2024against}.
Our contribution operationalizes discriminant validity as a statistical test, providing a concrete procedure that complements these conceptual frameworks.\looseness=-1

\paragraph{Proxy Outcomes and Label Bias.}
A related literature examines the consequences of training and evaluating algorithms on proxy outcomes that may not reflect the true quantity of interest.
\citet{obermeyer2019dissecting} show that a healthcare algorithm trained on healthcare costs as a proxy for health needs systematically underestimated the needs of Black patients.
\citet{fogliato2021validity} question whether arrest is a valid proxy for offense in recidivism prediction due to differential validity by race.
Causal frameworks analyze how proxy labels affect algorithmic decisions \citep{guerdan2023ground}.
These papers highlight the need for evaluation methodologies tailored to address proxy outcomes, motivating our falsification framework.\looseness=-1

\paragraph{Algorithmic Fairness Evaluation.}
Algorithmic fairness typically compares performance metrics across sensitive groups \citep{barocas2023fairness, chouldechova2018frontiers, mehrabi2021survey}.
While essential, these approaches do not directly address whether an algorithm is predicting its intended construct.
An algorithm can satisfy group fairness criteria while still fundamentally predicting the wrong quantity \citep{jacobs2021measurement,coston2023validity}.
Our falsification framework complements fairness evaluation by providing a validity check.
When our test fails, it suggests that foundational issues in problem formulation need to be addressed, regardless of the result of fairness evaluation.\looseness=-1

 \paragraph{Adversarial fairness.} Adversarial fairness methods learn representations that minimize loss on the prediction target while maximizing loss on a sensitive attribute \citep{edwards2016censoring, madras2018learning, zhang2018mitigating}.
Our falsification method similarly compares losses across variables, but differs in three key respects: (1)~our goal is \emph{relative}---we require that loss with respect to the target be statistically less than loss with respect to the sensitive attribute, rather than striving to maximize sensitive attribute loss; (2)~we compare losses after outcome-specific calibration, ensuring differences reflect predictive performance disparities rather than artifacts of base rates; and (3)~our method evaluates a fixed model rather than learning one anew.

 \section{Falsification as an Evaluation Strategy}
\label{sec:falsification in other domains}

In this section, we describe the logic of falsification and explore the well-developed practice of falsification in several domains.
This exposition lays the groundwork for our falsification procedure for predictive algorithms in Section~\ref{sec:method}. 

\subsection{Philosophical Logic of Falsification}

Scientific claims are typically structured so that they can be falsified by evidence, and scientific practice then subjects claims to tests that could disprove them.
In the Popperian tradition, empirical tests are designed to expose violations of a proposed property \citep{popper1959logic,popper1963conjectures}.
Deborah Mayo formalizes this idea in statistical terms through \emph{severe testing}: a claim gains credibility by passing a test that would, with high probability, have detected a flaw if one existed \citep{mayo2018statistical}.
These perspectives motivate the design of empirical procedures that prioritize detecting violations of a proposed property.
Sec~\ref{sec:method} proposes a falsification procedure that follows this logic: We treat the claim that the algorithm predicts the intended construct as a scientific hypothesis that must survive attempts at falsification.
Accordingly, our test is constructed so that models violating discriminant validity will, with high probability, be identified as such. 

\subsection{Falsification in causal inference: \emph{If the effect is real, it shouldn’t appear where it logically can’t}}

Falsification is often used to address a key challenge in causal inference: differentiating true effects from spurious ones — ``effects'' arising merely due to some artifact of the study.
In a randomized trial of a headache pill, observed pain reduction could reflect the pill's true effect or merely the ``effect of suggestion'' \citep{beecher1955powerful}.
Placebo treatments address this problem by introducing an inert ``placebo'' treatment.
Key to the placebo design is falsification: we designate a scenario (the placebo-controlled group) where we should not observe the effect, and if we do indeed observe the ``effect’’ here, then that calls into doubt whether the observed effect represents what we believe it does.
Inspired by this, in our falsification framework we identify a setting where the signal should not appear (predicting an impermissible proxy) and we test whether it does.\looseness=-1

Placebo tests are also used in the social sciences to assess credibility of findings.
A commonly used placebo test in observational studies is the \emph{negative control}, an auxiliary analysis where the researcher checks for an association that should be absent if the assumptions underlying the design hold but might be present if those assumptions are violated in some relevant way.
The negative control typically applies the main analysis method to an outcome or setting where we expect no effect to occur.
For example, studies that look at policy impacts will often ``test'' for the effect before the policy was enacted; if effects appear before the law was enacted, this suggests confounding rather than a causal effect \citep{card1994minimum, autor2003rise}.\looseness=-1

In our setting, the impermissible proxy plays a role analogous to a placebo or negative control: it defines a setting in which strong predictive performance would call into question the validity of the model. 
Predicting an impermissible proxy as well as the intended outcome constitutes grounds to falsify the claim that the model predicts the intended construct.\looseness=-1

\subsection{Falsification in Measurement}
In disciplines that develop tools to measure underlying constructs, such as psychometrics, falsification is important for establishing discriminant validity---the requirement that a measure does not capture unintended constructs \citep{campbell1959convergent}. 
Discriminant validity is a key component of construct validity, which more broadly concerns whether the tool measures the intended construct \citep{cronbach1955construct}. For example, a tool intended to measure depression should show discriminant validity with respect to anxiety. Depression and anxiety often co-occur, so it is possible for a tool to appear to measure anxiety when it may actually better measure depression \citep{hewitt1991beck}. For proper clinical use, it is important to establish discriminant validity to ensure appropriate diagnosis and treatment.

Among the methods used to assess discriminant validity in measurement domains, particularly relevant to our prediction setting is \emph{construct-based differential prediction}.\footnote{Other common methods for assessing discriminant validity require multiple measures of the underlying constructs. The \emph{multitrait-multimethod (MTMM) matrix} \citep{campbell1959convergent} assesses discriminant validity by comparing whether correlations between different traits measured by the same method are lower than correlations between the same trait measured by different methods. 
\emph{Factor analysis} evaluates whether items designed to measure different constructs load onto distinct factors and exhibit low cross-loadings, with good model fit and less-than-perfect factor correlations supporting discriminant validity \citep{brown2015cfa}. 
These approaches rely on redundancy in measurement, which is not available in our setting where we typically have only one measurement for the impermissible proxy.} This approach evaluates whether a measure predicts outcomes related to construct \(A\) better than outcomes related to construct \(B\). If a tool intended to measure \(A\) predicts outcomes associated with \(B\) equally well, this suggests a failure of discriminant validity \citep{darlington1971another,cohen2003applied}. 
Traditional differential prediction methods compare correlations across constructs. This approach is descriptive, offering no principled threshold for determining what correlation values constitute a meaningful violation of discriminant validity \citep{bollen1989sem,westen2003quantifying}. \looseness=-1

Our framework builds on this paradigm by adopting the logic of severe testing to formalize this as a hypothesis testing problem. 
Rather than descriptively comparing associations, we define a concrete failure mode—\emph{loss indiscriminance} (Def.~\ref{def:loss indiscriminance})—and construct a hypothesis test that is likely to detect this failure when it occurs. 

\section{Problem Formulation}
\label{sec:problem_formulation}

We observe a fixed predictive algorithm $\hat f: \mathcal{X} \rightarrow [0,1]$ and evaluation dataset $\mathcal{D} = \{(x_i, y_{i,1}, \ldots, y_{i,M}, \tilde{y}_i)\}_{i=1}^n$ where $X \in \mathcal{X}$ denotes features; $Y_1,\ldots,Y_M$ denotes $M \geq 1$ binary proxy outcomes; and $\tilde{Y}$ denotes a binary random variable that is impermissible as a target of prediction due to ethical or functional considerations. 
The ideal prediction target, $Y^{*}$, is unobservable.
Instead, we observe proxy outcomes $Y_1,\ldots,Y_M$ that are each a reasonable proxy for the intended outcome -- that is, each $Y_j$ is considered a \textbf{permissible proxy}.
The \textbf{impermissible proxy} $\tilde{Y}$ may describe a sensitive attribute like race or gender or may describe a different outcome than the intended target.

Our goal is to evaluate $\hat f$ using $\mathcal{D}$ and a loss function $\ell:\{0,1\}\times[0,1]\to\mathbb{R}_{\ge 0}$.
A common approach in practice when we lack access to $Y^{*}$ is to select a permissible proxy $Y_j$ and treat it as if it were $Y^{*}$, estimating $E[\ell(Y_j, \hat{f}(X))]$.
Unfortunately such an approach can be brittle, depending heavily on the choice of proxy outcome \citep{obermeyer2019dissecting,watson2023multi}.
Additionally, it's unclear what the bar for acceptable performance should be with respect to a proxy.
Most concerningly, such an approach risks greenlighting an algorithm that may be highly predictive of an impermissible proxy \citep{dastin2018amazon}.\looseness=-1 

\subsection{Falsifying Discriminant Validity of Predictive Algorithms}
\label{sec:falsify}
Inspired by the practice of falsification in other domains (see Section~\ref{sec:falsification in other domains}), we propose that a key step in evaluating algorithms involves assessing \emph{whether the algorithm predicts the impermissible proxy}.
At first glance, this may appear to be an easy question to answer, by analyzing performance metrics with respect to $\tilde{Y}$.
Because $\tilde{Y}$ may correlate with the unobservable outcome $Y^{*}$, the algorithm may exhibit some performance with respect to $\tilde{Y}$, but how to differentiate a reasonable modest level from a problematic level?
This requires specifying a threshold for “high” performance—ideally informed by domain knowledge and set prior to model development. 
Recognizing that such thresholds are often difficult to specify ex ante, we propose a data-driven method to answer the question \emph{Does the algorithm predict the impermissible proxy?}

To formulate the problem of falsifying predictive algorithms, we introduce a couple of terms. 
We define the random variable $L_Y := \ell(Y, \hat{f}(X))$ and the instance realization $L_{i,Y}:=\ell(y_i,\hat f(x_i))$. %
If we knew the population expected losses $\{E[L_{Y_j}]\}_{j=1}^M$ and $E[L_{\tilde Y}]$, then we could determine whether $E[L_{\tilde Y}] > E[L_{Y_j}]$ for all $j = 1, \ldots, M$ .
When this condition holds, we say there is \emph{loss discriminance}.

\begingroup
\setlength{\topsep}{3pt}
\setlength{\partopsep}{0pt}
\setlength{\parskip}{0pt}
\begin{definition}\label{def:loss discriminance}
\emph{Loss discriminance} holds when $E[L_{\tilde Y}] > E[L_{Y_j}]$ for all $j = 1, \ldots, M$.
\end{definition}
\begin{definition}\label{def:loss indiscriminance}
\emph{Loss indiscriminance} holds when $E[L_{\tilde Y}] \leq E[L_{Y_j}]$ for at least one $j = 1, \ldots, M$.
\end{definition}
\begin{definition}(Falsification procedure) \label{def: falsification procedure}
Given an evaluation dataset $\mathcal{D}$ and prediction function $\hat{f}$, a \emph{falsification procedure} returns $\texttt{INDISCRIMINANT}$ with high probability under loss indiscriminance.
\end{definition}
\endgroup

A falsification procedure tells us that $E[L_{\tilde Y}]$ is indistinguishable from $\{E[L_{Y_j}]\}_{j=1}^M$ with high probability in the scenario where the expected losses are truly exchangeable.
We designate this as $\texttt{INDISCRIMINANT}$ following the terminology of construct validity. 
Our definition of a falsification procedure reflects the Popperian logic of falsification, in which empirical procedures are constructed to detect violations of a proposed property rather than to definitively certify that the property holds \citep{popper1959logic,popper1963conjectures}.
Our specific failure mode is loss indiscriminance. 
We guard against falsely establishing discriminant validity by ensuring that models exhibiting loss indiscriminance are unlikely to receive a DISCRIMINANT result. 
When the procedure does return DISCRIMINANT, this provides evidence that expected loss is lower for permissible outcomes than for the impermissible proxy. 
However, a failure to obtain this result is inconclusive, as the procedure is not guaranteed to detect discriminant validity in all settings.\looseness=-1

This formulation is designed for settings where it is hard to verify ground truth for $Y^{*}$. Our formulation does not require observing $Y^{*}$.
While the procedure cannot confirm that a model accurately captures $Y^{*}$, it can provide evidence against certain failure modes by ruling out alignment with impermissible alternatives. 
As long as the permissible proxies better reflect $Y^{*}$ than $\tilde{Y}$, our method can be used to assess discriminant validity.\looseness=-1

The presence of correlation between $\tilde{Y}$ and $Y^*$ does not, by itself, constitute a failure of discriminant validity. 
$\tilde{Y}$ and $Y^*$ may be highly correlated due to historical and structural inequities. 
In such settings, some predictive performance with respect to $\tilde{Y}$ is expected even for a valid model. 
Our framework distinguishes between benign correlation and problematic behavior by adopting a \emph{relative} criterion.
A failure arises not because the model predicts $\tilde{Y}$ at all, but because it predicts $\tilde{Y}$ as well as or better than the permissible proxies, indicating that the model’s predictive signal is more aligned with the impermissible outcome than with the intended construct.

\section{Method}
\label{sec:method}

In this section we develop a falsification procedure to evaluate the discriminant validity of probabilistic binary classifiers used as predictive algorithms.
We use a hypothesis testing framework to implement falsification logic.
Drawing on nonparametric hypothesis testing methods, our procedure makes minimal assumptions on the data-generating process.

Before introducing our procedure, we connect the hypothesis testing setup to the logic of falsification. 
Under the logic of falsification, a useful test is one that detects a specific way a claim could fail, and a claim gains support only by passing such a severe test \citep{popper1959logic,popper1963conjectures,mayo2018statistical}.
In our setting, the property under scrutiny is discriminant validity of the model with respect to $\tilde Y$.
The concrete failure mode is loss indiscriminance (Def.~\ref{def:loss indiscriminance}), which serves as our null hypothesis $H_0$. The alternate hypothesis $H_1$ is loss discriminance (Def.~\ref{def:loss discriminance}).
The falsification procedure (Def.~\ref{def: falsification procedure}) acts as the severe test: rejecting $H_0$ returns \texttt{DISCRIMINANT}, indicating evidence against the failure mode. Failing to reject (an \texttt{INDISCRIMINANT} result) means that we are unable to falsify the failure mode.
Type~I error ($\alpha$) corresponds to the probability of incorrectly declaring \texttt{DISCRIMINANT} when loss indiscriminance holds, and Type~II error ($\beta$) to the probability of missing true discriminant validity.
Our procedure is asymmetric by design.
We explicitly control the chance of a false \texttt{DISCRIMINANT} claim (Type I error), but we do not guarantee low Type II error.
Accordingly, \texttt{DISCRIMINANT} is evidence against the failure mode, whereas \texttt{INDISCRIMINANT} should be interpreted as ``not yet falsified'' rather than ``invalidated.'' 
In the terminology of Mayo's severe testing framework, our procedure is a severe test of discriminant validity: if the algorithm were in fact predicting the impermissible outcome as well as or better than permissible outcomes, the test would likely detect this violation \citep{mayo2018statistical}. \looseness=-1

In what follows, our method uses the function $rank: \mathbb{R}^d \to \mathbb{Z}_{>0}^d$, which maps a vector of real numbers to the ranks of its elements, and the sign function $\text{sign}(z) = +1$ if $z > 0$ and $-1$ if $z < 0$. We first consider the single-proxy setting before proposing a general-purpose procedure for settings with multiple permissible proxies.

\subsection{Falsification with a Single Permissible Proxy}
Suppose we have a single permissible proxy, denoted by  $Y_1$, in addition to our impermissible proxy $\tilde{Y}$.
In this case, our question is simple: does the algorithm predict $Y_1$ better than $\tilde{Y}$?
This question is naturally suited for hypothesis testing.

We propose a one-sided hypothesis test (See Alg.~\ref{alg:single_proxy}) of the null hypothesis that $E[L_{\tilde Y}] \le E[L_{Y_1}]$.
More formally, let $\Delta_i = L_{i,\tilde{Y}} - L_{i,Y_1}$ denote the difference in loss between the impermissible and permissible proxies for sample $i$.
Our null hypothesis is $H_0: \mathbb{E}[\Delta] \leq 0$.
The alternative hypothesis is $H_1: \mathbb{E}[\Delta] > 0$, which states that the algorithm predicts the permissible proxy better than the impermissible one.
Rejecting the null hypothesis provides evidence that the algorithm predicts the permissible proxy better than the impermissible one, while failing to reject suggests the algorithm may be indiscriminant.\looseness=-1

Importantly, before computing losses, we apply \emph{Platt scaling} \citep{platt1999probabilistic} separately to calibrate the algorithm’s predictions for each outcome variable on a randomly held-out calibration set.
This outcome-specific calibration ensures that loss comparisons between different proxies are made on a common, calibrated scale.
Without it, differences in loss may reflect differential miscalibration---due to differences in base rates or outcome coding---rather than genuine differences in predictive performance, potentially leading to incorrect conclusions about validity.
By calibrating each outcome separately, we ensure that each loss reflects the best achievable predictive performance given the model’s score, so that comparisons capture the recoverable signal about each outcome rather than artifacts of coding or base rate differences.

After obtaining the sample-specific differences in calibrated losses $\Delta_i$, we perform a test to assess whether the loss difference is statistically significant.
Alg.~\ref{alg:single_proxy} proposes using Wilcoxon signed-rank because it makes no parametric assumptions on the distribution of loss differences.
In practice, if the loss differences satisfy normality assumptions and are free of outliers, a paired t-test may be used instead to achieve higher statistical power. 

\begin{algorithm}
\caption{Falsification Procedure with Single Permissible Proxy}
\label{alg:single_proxy}
\small
\begin{algorithmic}[1]
\REQUIRE Evaluation dataset $\mathcal{D} = \{(\hat f(x_i), y_{i,1}, \tilde{y}_i)\}_{i=1}^n$, significance level $\alpha$
\ENSURE $\texttt{DISCRIMINANT}$ or $\texttt{INDISCRIMINANT (inconclusive)}$ 
\FOR{$y \in \{\tilde{Y}, Y_1\}$}
    \STATE Apply Platt scaling to obtain calibrated probabilities $\hat p_i^{y}$ on held-out calibration set
\ENDFOR
\STATE Compute losses: $L_{i,\tilde{Y}} = \ell(\tilde{y}_i, \hat p_i^{\tilde{Y}})$ and $L_{i,Y_1} = \ell(y_{i,1}, \hat p_i^{Y_1})$ for $i = 1, \ldots, n$
\STATE Compute loss differences: $\Delta_i = L_{i,\tilde{Y}} - L_{i,Y_1}$ for $i = 1, \ldots, n$
\STATE \textbf{Wilcoxon signed-rank test:} $H_0: \theta \leq 0$ vs $H_1: \theta > 0$ where $\theta$ is the location parameter of the loss differences
\STATE \quad Compute $W = \sum_{i=1}^n \text{sign}(\Delta_i) \cdot rank(|\Delta_i|)$ and p-value from $W$, $n$
\IF{p-value $\leq \alpha$}
    \RETURN $\texttt{DISCRIMINANT}$
\ELSE
    \RETURN $\texttt{INDISCRIMINANT (inconclusive)}$
\ENDIF
\end{algorithmic}
\end{algorithm}

 \textbf{Interpretation:} Our method is a falsification procedure that with probability $1-\alpha$ returns \texttt{INDISCRIMINANT} in the scenario where loss indiscriminance holds.
A \texttt{DISCRIMINANT} result provides evidence that the algorithm is \emph{not} predicting the impermissible quantity and in that sense, passes our falsification test.
On the other hand, the result \texttt{INDISCRIMINANT} is an inconclusive result in that it may indicate that the algorithm truly fails our test or it may reflect that our sample didn't have enough power to observe the difference.
At first glance one might think that our definition of Falsification Procedure (\ref{def: falsification procedure}) means that an \texttt{INDISCRIMINANT} result gives us high confidence that loss indiscriminance holds, but this is actually an example of an \emph{inverse fallacy}: Our Falsification Procedure is a guarantee about $P(\text{return INDISCRIMINANT} \mid \text{loss indiscriminance})$ whereas our interpretation of an \texttt{INDISCRIMINANT} result involves $P(\text{loss indiscriminance} \mid \text{return INDISCRIMINANT})$ .


 \textbf{Choice of Loss Function:} Our falsification procedure requires specifying a loss function to compare prediction quality across outcomes.
Common choices include log loss (binary cross-entropy) and Brier score (squared error).
Log loss more heavily penalizes confident incorrect predictions, making it sensitive to cases where the model assigns high probability to the wrong outcome.
Brier score treats all errors more uniformly and is less sensitive to extreme predictions.
We recommend choosing the loss function based on domain considerations: if confident errors are costly, log loss may be more appropriate; if robustness to extreme predictions is preferred, Brier score is better.
Our empirical analyses find that our qualitative conclusions are generally robust to the choice of loss function (see Appendix~\ref{app:ablation}).

 \subsection{Falsification with Multiple Permissible Proxies}
\label{subsec:multiple_proxy}

Now consider the setting where we have multiple permissible proxies $Y_1, \ldots, Y_M$ in addition to the impermissible proxy $\tilde{Y}$.
In this case, we want to assess whether the algorithm predicts the impermissible proxy better than \emph{any} of the permissible proxies.
This is a more complex question than the single proxy case, as it requires comparing the impermissible proxy against multiple permissible proxies simultaneously.

We propose a \emph{conditional rank test} (Algorithm~\ref{alg:multiple_proxy}) that compares the rank of $L_{i,\tilde Y}$ relative to $\{L_{i,Y_j}\}_{j=1}^M$ within each sample, where rank $=1$ denotes the lowest loss (best performance).
The test is based on a null hypothesis of \emph{loss exchangeability}: under loss indiscriminance, $L_{i,\tilde Y}$ should be exchangeable with $\{L_{i,Y_j}\}_{j=1}^M$, conditional on the set of losses for each sample.
As in the single proxy case, we first apply Platt scaling separately to calibrate the algorithm's predictions for each outcome variable.
For each sample $i$, we compute the loss with respect to each outcome using the calibrated predictions, resulting in $M+1$ losses.
We then rank these $M+1$ losses from smallest to largest within each sample.
Our test statistic is the mean of these row-wise ranks of the impermissible loss $L_{i,\tilde Y}$: $\bar{R} = \frac{1}{n}\sum_{i=1}^n R_i^{\tilde{Y}}$.

 We test the null hypothesis $H_0$ that the losses $L_{i,\tilde{Y}}, L_{i,Y_1}, \ldots, L_{i,Y_M}$ are exchangeable within each sample against the one-sided alternative $H_1$ that the impermissible proxy tends to incur higher loss (worse rank) than the permissible proxies.
In the absence of ties, exchangeability implies that ranks are uniformly distributed over $\{1, \ldots, M+1\}$, yielding $\mathbb{E}[\bar{R}] = \frac{M+2}{2}$; the permutation test remains valid regardless of ties.
$H_1$ corresponds to the algorithm predicting the permissible proxies better than the impermissible proxy and is therefore consistent with the algorithm passing our falsification test.

Algorithm~\ref{alg:multiple_proxy} describes a permutation test that directly operationalizes the exchangeability assumption by randomly permuting outcome labels within each sample and recomputing the test statistic.
This approach yields exact finite-sample validity (up to Monte Carlo error) and naturally accommodates ties in the ranked losses.
Alternatively, when ties are unlikely and computational efficiency is a priority, a normal approximation to the sampling distribution of $\bar{R}$ may be used.

\begin{algorithm}
    \caption{Falsification Procedure with Multiple Permissible Proxies}
    \label{alg:multiple_proxy}
    \small
    \begin{algorithmic}[1]
    \REQUIRE Evaluation dataset $\mathcal{D} = \{(\hat f(x_i), y_{i,1}, \ldots, y_{i,M}, \tilde{y}_i)\}_{i=1}^n$, significance level $\alpha$
    \ENSURE $\texttt{DISCRIMINANT}$ or $\texttt{INDISCRIMINANT (inconclusive)}$
    \FOR{$y \in \{\tilde{Y}, Y_1, \ldots, Y_M\}$}
        \STATE Apply Platt scaling to obtain calibrated probabilities $\hat p_i^{y}$ on held-out calibration set
    \ENDFOR
    \STATE Compute losses: $L_{i,\tilde{Y}} = \ell(\tilde{y}_i, \hat p_i^{\tilde{Y}})$ and $L_{i,Y_j} = \ell(y_{i,j}, \hat p_i^{Y_j})$ for $i = 1, \ldots, n$, $j = 1, \ldots, M$
    \STATE For each sample $i$, rank the $M\!+\!1$ losses and let $R_i^{\tilde{Y}}$ denote the rank of $L_{i,\tilde Y}$ (1 = lowest loss)
    \STATE Compute test statistic $\bar{R}_{obs} = \frac{1}{n}\sum_{i=1}^n R_i^{\tilde{Y}}$
    \STATE \textbf{Permutation test:} $H_0$: losses $L_{i,\tilde{Y}}, L_{i,Y_1}, \ldots, L_{i,Y_M}$ are exchangeable within each sample $i$ \quad vs \quad $H_1$: $R_i^{\tilde{Y}}$ tends to be large (impermissible proxy has higher loss)
    \STATE \quad For $b = 1, \ldots, B$: permute losses across the $M\!+\!1$ outcomes within each sample $i$; recompute $\bar{R}^{(b)}$
    \STATE \quad $p = \frac{1 + \sum_{b=1}^B \mathbf{1}\{\bar{R}^{(b)} \ge \bar{R}_{\text{obs}}\}}{B + 1}$
    \IF{p-value $\le \alpha$}
        \RETURN $\texttt{DISCRIMINANT}$
    \ELSE
        \RETURN $\texttt{INDISCRIMINANT (inconclusive)}$
    \ENDIF
    \end{algorithmic}
    \end{algorithm}

Our conditional rank test provides a natural extension of the single proxy case to multiple permissible proxies.
By ranking losses within each sample, we account for sample-specific variation in loss magnitudes while focusing on the relative performance of the impermissible proxy compared to the permissible proxies.
The test is nonparametric in the sense that it makes no assumptions about the distribution of losses.

As with the single proxy case, a \texttt{DISCRIMINANT} result provides evidence that the algorithm is \emph{not} predicting the impermissible quantity better than the permissible proxies, and in that sense, passes our falsification test.
An \texttt{INDISCRIMINANT} result is inconclusive and may indicate either that the algorithm truly fails our test or that our sample didn't have enough power to observe a difference. 


\addtocounter{subsection}{-1}
  \subsection{Mitigating an \texttt{INDISCRIMINANT} result} \label{section: mitigation}
When the test returns an \texttt{INDISCRIMINANT} result, the first recourse step should be to attempt to procure more evaluation data, especially if the existing evaluation data is limited in sample size.
Rerunning the falsification test for the same model on a larger evaluation dataset can provide insight into whether the initial \texttt{INDISCRIMINANT} result stemmed from a failure of discriminant validity or from low power due to sample size.

An \texttt{INDISCRIMINANT} result that persists even with sufficiently large evaluation data leaves open the question whether the issue lies with the prediction task itself (e.g., a problematic underlying construct) or with the proxy outcome used during training, in which case selecting a more appropriate target may resolve the concern. 
To navigate this, we recommend first  critically re-examining the problem formulation for potential validity violations, including identifying misalignments between the feasible prediction target and the unobservable intended target that could threaten construct validity.
Several resources to scaffold this process include \mbox{
\citet{jacobs2021measurement,coston2023validity,kawakami2024situate}}\hskip0pt
.
After addressing validity problems identified during this process (for instance, by changing the prediction target to another proxy outcome) or if no validity problems were identified, the analyst may wish to train additional models, potentially using fairness mitigation strategies like adversarial learning to improve chances of passing the falsification test.
If a prediction model trained on a different outcome would pass our test for discriminant validity, this would suggest the prediction task may be sound but that certain learning targets are more suitable than others.
Importantly, to assess the discriminant validity of a new model, we must appropriately account for multiple testing, which we turn to in the next section.\looseness=-1

 \subsection{Falsification with Multiple Impermissible Proxies and/or Multiple Candidate Models}
We may want to assess the discriminant validity of multiple predictive models and/or across multiple impermissible proxies.
Without appropriate safeguards, repeated testing can inflate the probability of falsely concluding a model has discriminant validity. \looseness=-1

       We present Alg.~\ref{alg:holm} as a falsification procedure for the setting where we have multiple hypotheses -- either due to multiple candidate models and/or multiple impermissible proxies.
Alg.~\ref{alg:holm} uses Holm's correction to control the family-wise error rate at level $\alpha$.
Holm's procedure is a more powerful alternative to the well-known Bonferroni correction.
Holm's correction involves ordering p-values from smallest to largest and sequentially comparing them to progressively less stringent significance thresholds, rejecting hypotheses until the first non-rejection occurs \mbox{
\citep{Holm1979}}\hskip0pt
 . 

Importantly, if one anticipates using fairness-aware models as mitigation if the original model fails to pass, one should add these models to the set of candidate models \emph{before} beginning any falsification tests.
We provide an example of this in Sec.~\ref{sec:empirical}.
We recognize that in practice, analysts may wish to perform sequential and adaptive falsification tests -- e.g., determining which mitigations to run \emph{after} observing initial results.
Recent advances in inferential methods under data-dependent hypothesis selection \citep{ramdas2023game, waudbysmith2024estimating, fithian2022conditional} provide frameworks for valid inference under such adaptivity. 
Extending our approach to this setting is a promising direction for future work.
Finally, we caution against data-dependent or exploratory testing strategies that would be vulnerable to p-hacking, such as repeatedly modifying model classes, feature sets, or proxy definitions until a desired result is obtained.
More broadly, failure to pass falsification checks should motivate careful reconsideration of problem formulation and proxy choice—guided by domain knowledge—rather than iterative model tuning driven by apparent statistical significance.\looseness=-1

\begin{algorithm}
\caption{Falsification Procedure with Multiple Hypotheses (Holm's Correction)}
\label{alg:holm}
\small
\begin{algorithmic}[1]
\REQUIRE $K \geq 1$ impermissible proxies $\{\tilde{Y}_1, \ldots, \tilde{Y}_K\}$, $F \geq 1$ candidate models $\{\hat{f}_1, \ldots, \hat{f}_F\}$, evaluation dataset $\mathcal{D}$, family-wise significance level $\alpha$
\ENSURE For each $(k, f) \in \{1,\ldots,K\} \times \{1,\ldots,F\}$: $\texttt{DISCRIMINANT}$ or $\texttt{INDISCRIMINANT}$
\STATE Let $\nu = K \times F$
\FOR{$k = 1, \ldots, K$, $f = 1, \ldots, F$}
    \STATE Apply Alg.~\ref{alg:single_proxy} or Alg.~\ref{alg:multiple_proxy} to test $\hat{f}_f$ against $\tilde{Y}_k$; record p-value $p_{k,f}$
\ENDFOR
\STATE Order p-values: $p_{(1)} \leq \cdots \leq p_{(\nu)}$; let $H_{(1)}, \ldots, H_{(\nu)}$ be the reordered hypotheses
\FOR{$j = 1, \ldots, \nu$}
    \IF{$p_{(j)} > \alpha / (\nu - j + 1)$}
        \STATE Mark $H_{(j)}, \ldots, H_{(\nu)}$ as $\texttt{INDISCRIMINANT (inconclusive)}$; \textbf{break}
    \ELSE
        \STATE Mark $H_{(j)}$ as $\texttt{DISCRIMINANT}$
    \ENDIF
\ENDFOR
\end{algorithmic}
\end{algorithm}

\section{Empirical Analysis}
\label{sec:empirical}

In this section we investigate our falsification procedure in two domains: an admissions setting and a recidivism prediction setting.
We apply the multiple-hypothesis procedure (Alg.~\ref{alg:holm}) with an error budget of $\alpha = 0.05$ in each domain.
Code is available here: \href{https://github.com/mandycoston/falsifying-discriminant-validity}{https://github.com/mandycoston/falsifying-discriminant-validity}.

\subsection{Evaluating a Law School Admissions Algorithm}

Suppose a law school admissions office has developed an algorithm to predict which applicants are most likely to succeed in law school.
Potential proxies for success include first-year GPA, GPA at graduation, and bar exam passage.
The admissions office wants to ensure that the algorithm is not simply admitting students of a particular gender or race.

We use the Law School Admission Council (LSAC) dataset,\footnote{We use the preprocessed version available at \url{https://github.com/damtharvey/law-school-dataset}, which includes first-year GPA as a proxy outcome---a variable missing from other commonly used LSAC datasets.
Note that this preprocessing has resulted in some missing observations, particularly among non-white applicants.} which contains information about law school applicants including LSAT scores, undergraduate GPA, first-year law school GPA, bar passage outcomes, and demographic information.
We split the data into training (50\%), calibration (25\%), and evaluation (25\%) sets.
All falsification tests and performance evaluations are conducted on the evaluation set.
Race is a binary variable indicating white (1) versus non-white (0), following the coding in the available dataset, with 93\% of the data having race designated as white.

We consider three candidate models, all trained to predict first-year GPA (above/below median) using LSAT scores and undergraduate GPA as features.
The first model, $\hat{f}_1$, is a standard logistic regression.
The second and third models use LAFTR (Learning Adversarially Fair and Transferable Representations) \citep{madras2018learning}, a fairness-aware method that aims to learn representations from which an adversary cannot predict the sensitive attribute.
We train $\hat{f}_2$ with an adversary targeting race and $\hat{f}_3$ with an adversary targeting gender.
For each LAFTR model, we test the corresponding impermissible proxy---the one the adversary was designed to remove---to assess whether the fairness intervention successfully establishes discriminant validity.
Architectural details are provided in Appendix~\ref{app:laftr_metrics}.


\textbf{Hypotheses: }
We have three candidate models and two impermissible proxies: $\tilde{Y}_1=\text{gender}$ and $\tilde{Y}_2=\text{race}$.
We test discriminant validity for each model as follows:
$H_1$: $\hat{f}_1$ predicts gender at least as well as the permissible proxies (first-year GPA, cumulative GPA, and bar passage);
$H_2$: $\hat{f}_1$ predicts race at least as well as the permissible proxies;
$H_3$: $\hat{f}_2$ (LAFTR, race adversary) predicts race at least as well as the permissible proxies;
$H_4$: $\hat{f}_3$ (LAFTR, gender adversary) predicts gender at least as well as the permissible proxies.

\paragraph{Correlation analysis:}
Before presenting our falsification results, we consider whether a common discriminant validity approach---correlation analysis---could answer our question.
Table~\ref{tab:correlation_table} reports Pearson correlations between model predictions and each proxy.\footnote{Results remain substantively unchanged for Spearman's $\rho$ and Kendall's $\tau$; see Table~\ref{tab:correlation_table_nonparametric} in Appendix~\ref{app:correlation_nonparametric}.}
The correlations for permissible proxies are larger than correlation with gender for all models, but the correlations for permissible proxies are only larger than the correlation with race for LAFTR(adv=race). 
Correlation analysis does not constitute a severe test of discriminant validity: it defines no concrete failure mode, provides no control of error rates, and operates on raw model scores without calibration—meaning that observed differences may reflect base rate disparities rather than genuine differences in predictive signal.
If we ignore these limitations and proceed to interpretation, we might be tempted to conclude that LAFTR(adv=race) has discriminant validity with respect to race.
However, as we see below, the results of our falsification test cast doubt on this conclusion.
\begin{table}[!htb]
\centering
\caption{Pearson correlation between model predictions and each proxy on the LSAC evaluation set.
All models are trained on LSAT and undergraduate GPA to predict first-year GPA.
The correlations of model predictions with permissible proxies are larger than correlation with \textcolor{red!50!orange}{gender}  for all models, but the correlations for permissible proxies are only larger than the correlation with \textcolor{red!80!black}{race} for LAFTR(adv=race). This result might appear to suggest that LAFTR(adv=race) has discriminant validity with respect to \textcolor{red!80!black}{race}, but our falsification procedure fails to reject the null for \emph{any} model including LAFTR(adv=race). Sec.~\ref{sec: falsification results LSAC } provides our falsification results.}
\label{tab:correlation_table}
\small
\begin{tabular}{lccccc}
\toprule
\textbf{Model} & \textbf{1styearGPA} & \textbf{GPA} & \textbf{Pass Bar} & \textcolor{red!80!black}{\textbf{Race}} & \textcolor{red!50!orange}{\textbf{Gender}} \\
\midrule
Logistic regression & 0.2241 & 0.2713 & 0.2945 & \textcolor{red!80!black}{0.3920} & \textcolor{red!50!orange}{0.0042} \\
LAFTR (adv=race) & 0.1633 & 0.2081 & 0.1367 & \textcolor{red!80!black}{0.0988} & \textcolor{red!50!orange}{$-$0.0117} \\
LAFTR (adv=gender) & 0.2236 & 0.2742 & 0.3106 & \textcolor{red!80!black}{0.4325} & \textcolor{red!50!orange}{$-$0.0031} \\
\bottomrule
\end{tabular}
\end{table}

\paragraph{Baseline Approach}
 We also consider the standard evaluation approach as a baseline.
This approach compares performance metrics computed with respect to permissible proxies to those computed with respect to an impermissible proxy.
Table~\ref{tab:performance_metrics} reports such metrics for $\hat{f}_1$ (See Appendix~\ref{app:laftr_metrics} for the metrics for $\hat{f}_2$ and $\hat{f}_3$).
The leftmost columns present three widely used measures: AUC (area under the ROC curve), AU PR (area under the precision–recall curve), and mean squared error (MSE).
However, these metrics can be misleading because they aggregate performance across all possible thresholds, many of which are irrelevant in practice \citep{kwegyir2023misuseauc}.
To address this, we also include positive predictive value (PPV) metrics in the rightmost columns because PPV answers the operational question of interest: among the applicants selected, what proportion will succeed?\footnote{
We evaluate PPV at selection rates corresponding to the law school tier structure in the data, in which roughly 2\% of applicants in the data attended tier 1 schools, 10\% attended tier 1-2, 50\% tier 1-3, and 75\% attended tier 1-4.}
The PPVs and AUC/AU-PR suggest that the model predicts race better than any permissible outcome---including first-year GPA, the outcome it was trained on---while predicting gender no better than chance.
Interestingly, MSE tells a different story: MSE for race is larger than MSE for GPA and 1styearGPA.
Moreover, whether observed differences are statistically meaningful depends on the magnitude of differences and the sample size.
Informal comparisons of baseline metrics can be misleading because they do not account for sampling variability, and people are known to underweight sample size when interpreting statistical evidence \citep{tversky1971belief, griffin1992weighing}.
We caution against drawing conclusions from differences in baseline metrics alone.
This is particularly important in settings with highly correlated permissible and impermissible proxies where similar performance metrics may make it difficult to distinguish between benign correlation and meaningful violations of discriminant validity. 
Our falsification procedure is designed for precisely these settings, providing a principled inferential guarantee that baseline metrics lack.

\begin{table}[!htb]
\centering
\caption{Baseline evaluation of performance metrics on the LSAC dataset.
PPV at top $k\%$ measures the positive predictive value when selecting the top $k\%$ of samples ranked by model predictions.
Colors indicate impermissible proxies (\textcolor{red!80!black}{race} and \textcolor{red!50!orange}{gender}).
The model performs well on race (AUC = 0.89), raising concerns about validity.
}
\label{tab:performance_metrics}
\small
\begin{tabular}{l@{\hspace{0.5em}}ccccccc}
\toprule
\textbf{Proxy} & \textbf{AUC} & \textbf{AU PR} & \textbf{MSE} & \textbf{PPV Top 2\%} & \textbf{PPV Top 10\%} & \textbf{PPV Top 50\%} & \textbf{PPV Top 75\%} \\
\midrule
1styearGPA & 0.6256 & 0.5993 & 0.2375 & 0.6489 & 0.6368 & 0.5901 & 0.5529 \\
GPA & 0.6547 & 0.6233 & 0.2320 & 0.6809 & 0.6603 & 0.6128 & 0.5660 \\
Pass Bar & 0.7629 & 0.9641 & 0.2477 & 0.9894 & 0.9915 & 0.9649 & 0.9466 \\
\textcolor{red!80!black}{Race} & \textcolor{red!80!black}{0.8948} & \textcolor{red!80!black}{0.9908} & \textcolor{red!80!black}{0.2461} & \textcolor{red!80!black}{1.0000} & \textcolor{red!80!black}{1.0000} & \textcolor{red!80!black}{0.9923} & \textcolor{red!80!black}{0.9854} \\
\textcolor{red!50!orange}{Gender} & \textcolor{red!50!orange}{0.5019} & \textcolor{red!50!orange}{0.5726} & \textcolor{red!50!orange}{0.2641} & \textcolor{red!50!orange}{0.6064} & \textcolor{red!50!orange}{0.5897} & \textcolor{red!50!orange}{0.5605} & \textcolor{red!50!orange}{0.5680} \\
\bottomrule
\end{tabular}
\end{table}

\subsubsection{Falsification Test Results} \label{sec: falsification results LSAC }
We apply Alg.~\ref{alg:multiple_proxy} to each hypothesis, using cumulative GPA, first-year GPA, and bar passage as the permissible proxies.

\begin{figure}[h]
\centering
\includegraphics[width=0.75\textwidth]{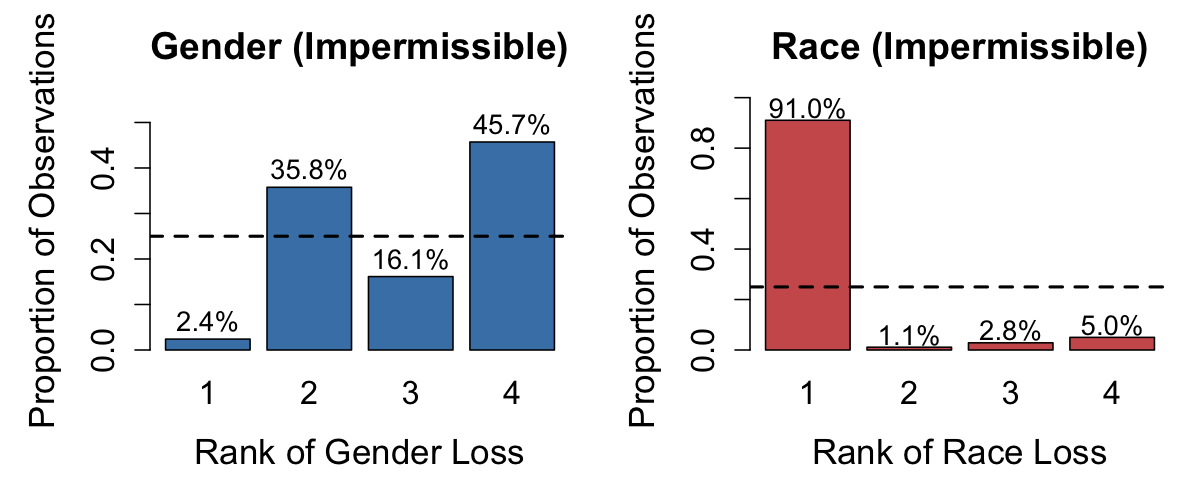}
\caption{Rank distribution of impermissible proxies (gender and race) for Alg. 2 applied to the logistic regression ($\hat{f}_1$) on the LSAC dataset.
For each observation, we rank the calibrated log-loss for the impermissible proxy against the three permissible proxies (cumulative GPA, first-year GPA, bar passage).
Rank 1 = best predicted; Rank 4 = worst predicted.
The dashed line shows the expected 25\% under the null hypothesis.
\textbf{Left:} Gender is worst-predicted (rank 4) for 45.7\% of observations, and the model passes our test ($p \approx 0$) indicating discriminant validity with respect to gender.
\textbf{Right:} Race is best-predicted (rank 1) for 91\% of observations, indicating the model predicts race better than permissible outcomes.
The model fails to pass our test of discriminant validity with respect to race ($p \approx 1$).}
\label{fig:rank_distribution}
\end{figure}

\paragraph{Logistic regression ($\hat{f}_1$)}
For gender ($H_1$), as shown in Fig.~\ref{fig:rank_distribution} (left), gender ranks as the worst-predicted proxy (rank 4) for 45.7\% of observations and as the best-predicted (rank 1) for only 2.4\%.
The raw p-value is $p_1 \approx 0$.
For race ($H_2$), as shown in Fig.~\ref{fig:rank_distribution} (right), the results are strikingly different: race ranks as the best-predicted proxy (rank 1) for 91\% of observations and as the worst (rank 4) for only 5\%.
The raw p-value is $p_2 \approx 1$.
This raises serious validity concerns because it suggests that a model intended to predict academic success is actually better at predicting race than at predicting proxies for success. 

\paragraph{LAFTR models ($\hat{f}_2$, $\hat{f}_3$)}
For $\hat{f}_2$ (LAFTR with race adversary) tested against race ($H_3$), the raw p-value is $p_3 \approx 1$.
For $\hat{f}_3$ (LAFTR with gender adversary) tested against gender ($H_4$), the raw p-value is $p_4 \approx 0$.

\paragraph{Inference via Holm's Procedure}
We have four hypotheses, so Holm's procedure uses thresholds $\alpha/(4-j+1)$ for the $j$-th ordered p-value: $\alpha_1 = 0.05/4 = 0.0125$, $\alpha_2 = 0.05/3 \approx 0.0167$, $\alpha_3 = 0.05/2 = 0.025$, and $\alpha_4 = 0.05$.
Our ordered p-values are $p_{(1)} = p_1 \approx 0$, $p_{(2)} = p_4 \approx 0$, $p_{(3)} = p_2 \approx 1$, and $p_{(4)} = p_3 \approx 1$.
Since $p_{(1)} < 0.0125$, we reject $H_1$ (logistic regression, gender).
Since $p_{(2)} < 0.0167$, we reject $H_4$ (LAFTR gender adversary, gender).
Since $p_{(3)} \approx 1 > 0.025$, we fail to reject $H_2$ and stop; $H_3$ is also not rejected.
We conclude that the models establish discriminant validity with respect to gender, but no model---including those with fairness-aware adversarial training---establishes discriminant validity with respect to race.

Interestingly, the LAFTR model does not pass our falsification test for race. 
In particular, the calibration step is key to understanding this finding: without Platt scaling, both $\hat{f}_1$ (logistic regression with no fairness adjustment) and $\hat{f}_2$ (LAFTR with race-adversary) would appear to pass the falsification test for race (p-values $\approx 0$), but after outcome-specific recalibration, they do not (p-values $\approx 1$; See Appendix~\ref{app:ablation} for ablation without Platt scaling).
Calibration is particularly important in settings like this where the base rate for the prediction target (0.5) differs significantly from the base rate of the impermissible proxy (0.93).
In such settings, the raw model predictions may be quite far from the impermissible outcome due to base rate differences, while still encoding information about the impermissible proxy that could be recovered via calibration.
This highlights the importance of using calibration to shift the comparison from the model’s raw outputs to the information they contain about each outcome.

A limitation of our analysis is that it does not account for decision dependencies within the permissible proxies.
For instance, a student’s GPA or bar passage outcome may depend on the school they attended.
Observed outcomes may differ from the actual predictive target: potential success at the specific school making the decision \citep{guerdan2023ground, coston2023validity}.
Our framework is compatible with counterfactual modeling that addresses decision dependencies \citep{rambachan2022robust, coston2020counterfactual}, but we forgo these methods here due to data constraints.
We urge analysts to consider these causal dependencies when formulating evaluation targets in practice.

\subsection{Falsification of the COMPAS Recidivism Prediction Instrument}
\label{sec:compas}
Next we explore the advantages and limitations of our method by applying it to the recidivism prediction setting.
The goal is to predict which defendants are likely to re-offend.
The recidivism prediction task is controversial for a variety of reasons, including the notable problem that we do not observe re-offense and must rely on proxies for re-offense such as re-arrest \citep{bao2021compasicated, jacobs2021measurement, wang2024against,johndrow2019algorithm}.
The use of such a proxy immediately invokes concerns of construct validity \citep{fogliato2021validity, jacobs2021measurement, coston2023validity,guerdan2023ground}.
In this section we show how our falsification test can be used to investigate one aspect of construct validity -- divergent validity.
Our method does not assess other key components of construct validity such as convergent validity, which itself is highly questionable in the recidivism prediction setting.
We explore this below, emphasizing that our method is not a comprehensive tool for evaluating validity; rather, it assesses a specific aspect of validity (divergent validity) and should be used alongside other approaches to evaluate other dimensions of validity. 

We present analysis on the COMPAS dataset, which contains information about defendants in the criminal justice system in Broward County, Florida, including demographic information, recidivism outcomes, and the COMPAS risk scores produced by a proprietary risk assessment instrument.
The COMPAS dataset is widely used and widely criticized in the fairness literature \citep{bao2021compasicated, fogliato2021validity}.
The criminal justice system has systemic biases against black defendants, affecting both who is represented in datasets like COMPAS and how their data is measured \citep{alexander2011new}.
\citet{bao2021compasicated} provide an in-depth discussion on problems with the COMPAS dataset.
They call attention to biases in how the data is collected and measured, arguing that failure to account for these structural limitations can render fairness analyses misleading. 

Prior work shows that recidivism models rely heavily on two predictive features, defendant's age and their prior offenses \citep{angwin2016machine,DresselFarid2018, rudin2019stop, flores2016}.
Researchers have shown that simple models that rely only on these features perform comparably to more complex models \citep{DresselFarid2018, rudin2019stop}.
This seminal work has deepened our understanding of the prediction task in the recidivism setting while also raising questions about the validity of predictive models.
For instance, can we establish that recidivism prediction models assess risk beyond simply flagging young defendants? 

Using the language of our falsification procedure, we ask the question \emph{whether recidivism models predict the re-arrest outcome better than the impermissible proxy (age $< 25$)}.
Under this framing, we treat re-arrest within two years of release as a permissible proxy for recidivism, while emphasizing that this designation is highly qualified.
Re-arrest is a problematic proxy for underlying criminal behavior for several well-documented reasons: many crimes go undetected, enforcement intensity varies across contexts, and certain demographic groups are more likely to be arrested even when holding underlying crime rates constant \citep{alexander2011new,fogliato2021validity}.
As such, the permissibility of re-arrest as a proxy is itself debatable.
The recidivism setting is notoriously difficult to obtain valid proxies for crime offenses.
We therefore proceed using re-arrest as an imperfect proxy solely for the purpose of illustrating our falsification procedure and with the explicit disclaimer that we are not advocating for the use of this proxy to build predictive models.
In addition to treating age as the impermissible proxy, we also conduct a test where race is the impermissible proxy.

\textbf{Hypotheses: }
We have one candidate model, denoted $\hat{f}_1$, namely the COMPAS decile risk score.\footnote{Decile risk scores are coarsened values of the raw COMPAS risk scores ranging from 1 to 10.
We do not use the raw risk scores because they are not available in the dataset, and moreover the raw risk scores are not used in practice for decision making.
Instead, typically a three-tier risk category (low, medium, and high) constructed from the decile risk scores is used \citep{angwin2016machine}.We obtain similar results when using the three-tier risk category as the model under evaluation.
We also obtain similar results when using raw scores of a predictive model that we train on re-arrest outcomes (See Appendix~\ref{app:compas_additional_results})} 
The impermissible proxies are $\tilde{Y}_1=\mathbf{1}\{\text{Age} < 25\}$ and $\tilde{Y}_2=\mathbf{1}\{\text{Race = Black}\}$.
The permissible outcome is two-year re-arrest.
The null hypotheses are $H_1$: $\hat{f}_1$ predicts $\mathbf{1}\{\text{Age} < 25\}$ at least as well as two-year re-arrest, and $H_2$: $\hat{f}_1$ predicts race at least as well as two-year re-arrest.

To assess whether the model predicts the impermissible proxy (age $< 25$) better than it predicts two-year re-arrest, we apply Alg. 1.
Our concern is not whether age is permissible to use in prediction, but whether the score is predicting a risk signal beyond age.
This yields a raw p-value for $H_1$: $p_1 \approx 1.00$.

\begin{table}[h]
\centering
\caption{Baseline evaluation of performance metrics for COMPAS decile scores.
TNR and PPV at top $k\%$ measure true negative rate and positive predictive value when flagging the top $k\%$ highest-risk individuals.
Colors indicate impermissible proxies (\textcolor{red!80!black}{race} and \textcolor{red!50!orange}{age}).
AUC with respect to race and age is slightly less than with respect to re-arrest, illustrating why standard metrics alone cannot assess discriminant validity.}
\label{tab:compas_metrics}
\small
\begin{tabular}{lccccccc}
\toprule
\textbf{Outcome} & \textbf{AUC} & \textbf{AU PR} & \textbf{MSE} & \textbf{TNR Top 10\%} & \textbf{TNR Top 25\%} & \textbf{PPV Top 10\%} & \textbf{PPV Top 25\%} \\
\midrule
Re-arrest & 0.7022 & 0.6427 & 0.2412 & 0.9394 & 0.8375 & 0.7306 & 0.6772 \\
\textcolor{red!50!orange}{Age $<$ 25} & \textcolor{red!50!orange}{0.6896} & \textcolor{red!50!orange}{0.3135} & \textcolor{red!50!orange}{0.2335} & \textcolor{red!50!orange}{0.8953} & \textcolor{red!50!orange}{0.7595} & \textcolor{red!50!orange}{0.3322} & \textcolor{red!50!orange}{0.3148} \\
\textcolor{red!80!black}{Race (Black)} & \textcolor{red!80!black}{0.6793} & \textcolor{red!80!black}{0.6675} & \textcolor{red!80!black}{0.2677} & \textcolor{red!80!black}{0.9360} & \textcolor{red!80!black}{0.8380} & \textcolor{red!80!black}{0.7475} & \textcolor{red!80!black}{0.7143} \\
\bottomrule
\end{tabular}
\end{table}

Next we investigate discriminant validity with respect to the defendant's race.
We run Alg.~1 with race as the impermissible proxy, yielding the raw p-value for $H_2$: $p_2 = 0.025504$.

\paragraph{Inference via Holm's Procedure}
The ordered p-values are $p_{(1)} = p_2= 0.025504$ and $p_{(2)} = p_1 \approx 1$.
Holm's first threshold is $\alpha_1 = 0.05/2 = 0.025$.
Because $0.025504 > 0.025$, we fail to reject $H_2$.
Under Holm's step-down procedure, we then stop immediately and fail to reject both hypotheses in the COMPAS family.
Thus, we fail to establish discriminant validity with respect to either age or race.
This finding casts doubt on the validity of COMPAS risk scores as models of signal beyond simply identifying young defendants.
However, because an \texttt{INDISCRIMINANT} result is inconclusive, we cannot definitively say anything beyond that there was insufficient evidence to establish discriminant validity.
Importantly, standard evaluation would be tempted to draw a stronger conclusion here, whereas our framework forces an explicit acknowledgment of uncertainty.

We briefly comment on the low p-value for $H_2$, which fell just shy of the border for rejecting $H_2$.
This tenuous result highlights an important limitation of our method. 
First, because our method only tests divergent validity, a predictive model could pass our method even when it was predicting a proxy that fails convergent validity.
For instance, re-arrest may be a poor proxy for crime, even if we can predict it better than we can predict the impermissible proxy.

We contrast our analysis to standard evaluation practices that compare performance metrics with respect to re-arrest against metrics with respect to age.
Table~\ref{tab:compas_metrics} reports such metrics.
Under this standard evaluation, our conclusions are highly sensitive to both metric choice and our interpretation of differences.
Re-arrest has only a slightly higher AUC (0.7022) than age (0.6896), making it unclear whether this is a meaningful difference.
While TNR and PPV suggest a larger gap favoring arrest, MSE (mean-squared error) indicates that re-arrest has slightly higher error (0.24) than age (0.23).
This highlights the challenge of assessing discriminant validity without a principled statistical test.
Our approach gives a statistically principled way to determine whether a model predicts an intended outcome better than an impermissible proxy. 

We conclude this empirical section with a note on external validity.
This analysis assesses discriminant validity on a biased subsample of defendants -- those who had the opportunity to re-offend.
This sample excludes defendants who were detained.
The purported use of recidivism prediction tools is to assess \emph{all} defendants -- including those who will not be released.
This selection bias calls into question the \emph{external validity} of our evaluation \citep{coston2023validity}.
Were such an evaluation to be used in real-world practice to inform the use of predictive models, it would be imperative to address this selection bias using methods such as imputation or counterfactual evaluation \citep{coston2020counterfactual,rambachan2022robust,kleinberg2018human}.

\begin{figure}[t]
    \centering
    \includegraphics[width=0.85\textwidth]{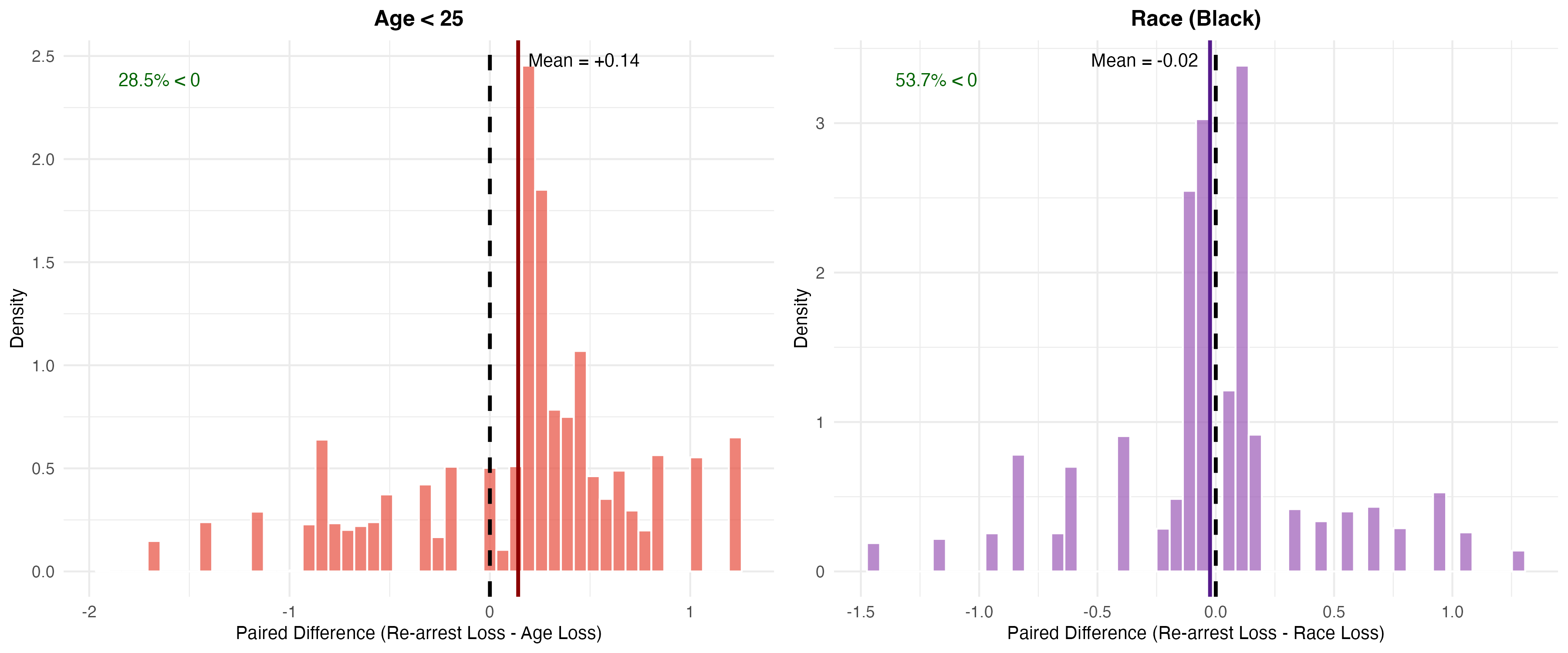}
    \caption{Paired difference distributions for Alg.~1 on COMPAS with age (left) and race (right) as impermissible proxies. Histograms show the distribution of (re-arrest loss $-$ impermissible proxy loss) across observations. Dashed black lines mark zero. \textbf{Left (Age):} Fail to establish discriminant validity \emph{wrt} age ($p \approx 1$). \textbf{Right (Race):} Borderline result for discriminant validity \emph{wrt} race ($p = 0.025$).}
    \label{fig:comparison}
    \end{figure}

\section{Conclusion}
\label{sec:conclusion}

We propose a falsification framework for predictive algorithms that addresses the fundamental question whether the algorithm predicts intended quantities better than impermissible ones.
Drawing on the well-established practice of falsification in other fields, our framework provides a principled statistical test for discriminant validity---the requirement that an algorithm not capture constructs it is not intended to measure.
We emphasize that our falsification approach should be one component of a broader evaluation strategy.
Passing our test does not imply that an algorithm is fair, appropriate, or ready for deployment.
Rather, our framework provides a necessary but not sufficient check on algorithmic validity.
Our method accommodates setting where we only have proxies for the intended but unobservable outcome, making it applicable to settings where the desired construct---such as health need, future job performance, or criminal behavior---is inherently difficult to define or impossible to measure directly.
In such settings, the test can still provide evidence about whether a model's predictive signal is more aligned with permissible proxies than with impermissible ones, even though it cannot verify alignment with the unobservable construct itself.
When an algorithm fails our test, the appropriate response is not to adjust model parameters but to return to problem formulation and scrutinize design choices, data limitations, and the suitability of the task for prediction.

\section*{Acknowledgments}
This project is supported by The AI2050 Program at Schmidt Sciences. We thank the anonymous reviewers for their valuable feedback.

\section*{Generative AI Statement}
During the preparation of this work, the author used generative AI to identify grammatical problems and to help with formatting.
The author also used generative AI to test code for correctness.
After obtaining suggestions from the genAI, the author reviewed and edited the changes as needed and take full responsibility for the content of the publication.
All analyses and conclusions are our own.

\section*{Ethical Considerations}
The empirical section of this paper involves discussion of decision making in consequential domains such as admissions and criminal justice.
The criminal justice system in particular is ethically fraught due to a history of systemic biases and harms perpetuated against Black people and other marginalized communities \citep{alexander2011new}.
As a result, analyses involving criminal justice data must be interpreted with care, recognizing that observed data reflect not only individual behavior but also patterns of policing, surveillance, and enforcement that are themselves shaped by structural inequities.

Our use of criminal justice data is intended solely to illustrate the methodological properties, strengths, and limitations of our falsification framework, not to endorse the use of predictive risk assessment tools in practice.
We emphasize that passing our falsification test is not sufficient to conclude that a model is fair, appropriate, or suitable for deployment -- such a conclusion must be based on a variety of sociotechnical evaluations and considerations.
We believe that proper use of validity-centered evaluation can serve as a harm-reducing practice: by identifying when algorithms predict unintended or impermissible quantities, such analyses can help prevent the legitimization and deployment of systems that risk reinforcing existing inequities.


\bibliographystyle{plainnat}
\bibliography{references}

@article{beecher1955powerful,
  title        = {The Powerful Placebo},
  author       = {Beecher, Henry K.},
  journal      = {Journal of the American Medical Association},
  volume       = {159},
  number       = {17},
  pages        = {1602--1606},
  year         = {1955},
  doi          = {10.1001/jama.1955.02960340022006}
}

@book{barocas2023fairness,
  title={Fairness and Machine Learning: Limitations and Opportunities},
  author={Barocas, Solon and Hardt, Moritz and Narayanan, Arvind},
  year={2023},
  publisher={MIT Press},
  address={Cambridge, MA},
  url={https://fairmlbook.org}
}

@inproceedings{mehrabi2021survey,
  title={A survey on bias and fairness in machine learning},
  author={Mehrabi, Ninareh and Morstatter, Fred and Saxena, Nripsuta and Lerman, Kristina and Galstyan, Aram},
  booktitle={ACM computing surveys (CSUR)},
  volume={54},
  number={6},
  pages={1--35},
  year={2021},
  organization={ACM}
}

@book{popper1959logic,
  title={The Logic of Scientific Discovery},
  author={Popper, Karl R.},
  year={1959},
  publisher={Hutchinson \& Co},
  address={London},
  note={Originally published in German as "Logik der Forschung" in 1934}
}

@book{popper1963conjectures,
  title={Conjectures and Refutations: The Growth of Scientific Knowledge},
  author={Popper, Karl R.},
  year={1963},
  publisher={Routledge and Kegan Paul},
  address={London}
}

@book{mayo2018statistical,
  title={Statistical Inference as Severe Testing: How to Get Beyond the Statistics Wars},
  author={Mayo, Deborah G.},
  year={2018},
  publisher={Cambridge University Press},
  address={Cambridge}
}

@book{barnett1994outliers,
  title={Outliers in statistical data},
  author={Barnett, Vic and Lewis, Toby and others},
  volume={3},
  number={1},
  year={1994},
  publisher={Wiley New York}
}

@book{d2017goodness,
  title={Goodness-of-fit-techniques},
  author={D'Agostino, Ralph B},
  year={2017},
  publisher={Routledge}
}

@article{shapiro1965analysis,
  title={An analysis of variance test for normality (complete samples)},
  author={Shapiro, Samuel Sanford and Wilk, Martin B},
  journal={Biometrika},
  volume={52},
  number={3-4},
  pages={591--611},
  year={1965},
  publisher={Oxford University Press}
}

@book{Freedman2007,
  title     = {Statistics},
  author    = {Freedman, David and Pisani, Robert and Purves, Roger},
  edition   = {4},
  publisher = {W. W. Norton \& Company},
  address   = {New York},
  year      = {2007}
}

@inproceedings{platt1999probabilistic,
  title={Probabilistic outputs for support vector machines and comparisons to regularized likelihood methods},
  author={Platt, John},
  booktitle={Advances in large margin classifiers},
  pages={61--74},
  year={1999},
  organization={MIT Press}
}

@inproceedings{kwegyir2023misuseauc,
  title     = {The Misuse of AUC: What High Impact Risk Assessment Gets Wrong},
  author    = {Kwegyir-Aggrey, Kweku and Gerchick, Marissa and Mohan, Malika and Horowitz, Aaron and Venkatasubramanian, Suresh},
  booktitle = {Proceedings of the 2023 ACM Conference on Fairness, Accountability, and Transparency (FAccT '23)},
  year      = {2023},
  publisher = {Association for Computing Machinery},
  address   = {New York, NY, USA},
  pages     = {1--14},
  doi       = {10.1145/3593013.3594100},
  url       = {https://doi.org/10.1145/3593013.3594100}
}

@article{bao2021compasicated,
  title   = {It's COMPASicated: The Messy Relationship between RAI Datasets and Algorithmic Fairness Benchmarks},
  author  = {Bao, Michelle and Zhou, Angela and Zottola, Samantha and Brubach, Brian and Desmarais, Sarah and Horowitz, Aaron and Lum, Kristian and Venkatasubramanian, Suresh},
  journal = {arXiv preprint arXiv:2106.05498},
  year    = {2021},
  url     = {https://arxiv.org/abs/2106.05498}
}

@article{rudin2019stop,
  title   = {Stop Explaining Black Box Machine Learning Models for High Stakes Decisions and Use Interpretable Models Instead},
  author  = {Rudin, Cynthia},
  journal = {Nature Machine Intelligence},
  volume  = {1},
  number  = {5},
  pages   = {206--215},
  year    = {2019},
  doi     = {10.1038/s42256-019-0048-x},
  url     = {https://www.nature.com/articles/s42256-019-0048-x}
}

@article{barocas2016big,
  title={Big data's disparate impact},
  author={Barocas, Solon and Selbst, Andrew D},
  journal={Calif. L. Rev.},
  volume={104},
  pages={671},
  year={2016},
  publisher={HeinOnline}
}

@misc{dastin2018amazon,
  author       = {Jeffrey Dastin},
  title        = {Amazon scraps secret {AI} recruiting tool that showed bias against women},
  howpublished = {Reuters},
  year         = {2018},
  month        = {October 10},
  url          = {https://www.reuters.com/article/world/insight-amazon-scraps-secret-ai-recruiting-tool-that-showed-bias-against-women-idUSKCN1MK0AG/},
  note         = {Accessed: 2025-10-08}
}

@article{lipton2018does,
  title={Does mitigating ML's impact disparity require treatment disparity?},
  author={Lipton, Zachary and McAuley, Julian and Chouldechova, Alexandra},
  journal={Advances in neural information processing systems},
  volume={31},
  year={2018}
}

@misc{booth2024guardian,
  author       = {Robert Booth},
  title        = {Revealed: bias found in {AI} system used to detect {UK} benefits fraud},
  howpublished = {The Guardian},
  year         = {2024},
  month        = {December 6},
  url          = {https://www.theguardian.com/society/2024/dec/06/revealed-bias-found-in-ai-system-used-to-detect-uk-benefits},
  note         = {Accessed: 2025-10-15}
}

@misc{booth2024dwp,
  author       = {Robert Booth},
  title        = {{DWP} algorithm wrongly flags 200,000 people for possible fraud and error},
  howpublished = {The Guardian},
  year         = {2024},
  month        = {June 23},
  url          = {https://www.theguardian.com/society/article/2024/jun/23/dwp-algorithm-wrongly-flags-200000-people-possible-fraud-error},
  note         = {Accessed: 2026-10-015}
}

@misc{ho2022oregon,
  author       = {Sally Ho and Garance Burke},
  title        = {{Oregon} dropping {AI} tool used in child abuse cases},
  howpublished = {Associated Press},
  year         = {2022},
  month        = {June 2},
  url          = {https://apnews.com/article/politics-technology-pennsylvania-child-abuse-1ea160dc5c2c203fdab456e3c2d97930},
  note         = {Accessed: 2026-01-08}
}

@misc{ho2023pittsburgh,
  author       = {Sally Ho and Garance Burke},
  title        = {Justice Department scrutinizes {Pittsburgh} child welfare {AI} tool},
  howpublished = {Associated Press},
  year         = {2023},
  month        = {January 10},
  url          = {https://apnews.com/article/justice-scrutinizes-pittsburgh-child-welfare-ai-tool-4f61f45bfc3245fd2556e886c2da988b},
  note         = {Accessed: 2026-01-08}
}

@article{campbell1959convergent,
  title={Convergent and discriminant validation by the multitrait-multimethod matrix},
  author={Campbell, Donald T and Fiske, Donald W},
  journal={Psychological bulletin},
  volume={56},
  number={2},
  pages={81--105},
  year={1959},
  publisher={American Psychological Association}
}

@article{cronbach1955construct,
  title={Construct validity in psychological tests},
  author={Cronbach, Lee J and Meehl, Paul E},
  journal={Psychological bulletin},
  volume={52},
  number={4},
  pages={281--302},
  year={1955},
  publisher={American Psychological Association}
}

@inproceedings{coston2023validity,
  title={A validity perspective on evaluating the justified use of data-driven decision-making algorithms},
  author={Coston, Amanda and Kawakami, Anna and Zhu, Haiyi and Holstein, Ken and Heidari, Hoda},
  booktitle={2023 IEEE conference on secure and trustworthy machine learning (SaTML)},
  pages={690--704},
  year={2023},
  organization={IEEE}
}

@article{freiesleben2025benchmarking,
  title={The benchmarking epistemology: Construct validity for evaluating machine learning models},
  author={Freiesleben, Timo and Zezulka, Sebastian},
  journal={arXiv preprint arXiv:2510.23191},
  year={2025}
}

@book{eubanks2018automating,
  title={Automating Inequality: How High-Tech Tools Profile, Police, and Punish the Poor},
  author={Eubanks, Virginia},
  year={2018},
  publisher={St. Martin's Press},
  address={New York}
}

@misc{angwin2016machine,
  author={Angwin, Julia and Larson, Jeff and Mattu, Surya and Kirchner, Lauren},
  title={Machine Bias},
  howpublished={ProPublica},
  year={2016},
  month={May 23},
  url={https://www.propublica.org/article/machine-bias-risk-assessments-in-criminal-sentencing},
  note={Accessed: 2026-01-08}
}

@article{obermeyer2019dissecting,
  title={Dissecting racial bias in an algorithm used to manage the health of populations},
  author={Obermeyer, Ziad and Powers, Brian and Vogeli, Christine and Mullainathan, Sendhil},
  journal={Science},
  volume={366},
  number={6464},
  pages={447--453},
  year={2019},
  publisher={American Association for the Advancement of Science},
  doi={10.1126/science.aax2342}
}

@article{vyas2020hidden,
  title={Hidden in plain sight---reconsidering the use of race correction in clinical algorithms},
  author={Vyas, Darshali A and Eisenstein, Leo G and Jones, David S},
  journal={New England Journal of Medicine},
  volume={383},
  number={9},
  pages={874--882},
  year={2020},
  publisher={Mass Medical Soc},
  doi={10.1056/NEJMms2004740}
}

@article{gilman2020welfare,
  title={Poverty Algorithms: A Poverty Lawyer's Guide to Fighting Automated Decision-Making Harming Low-Income Communities},
  author={Gilman, Michele E},
  journal={UCLA Law Review},
  volume={67},
  pages={1326--1400},
  year={2020}
}

@misc{charette2018unemployment,
  author={Charette, Robert N},
  title={Michigan's Automated Unemployment System Failures: A Case Study in AI Ethics},
  howpublished={IEEE Spectrum},
  year={2018},
  month={October},
  url={https://spectrum.ieee.org/michigans-automated-unemployment-system-failures-a-case-study-in-ai-ethics},
  note={Accessed: 2026-01-08}
}

@inproceedings{passi2019problem,
  title={Problem formulation and fairness},
  author={Passi, Samir and Barocas, Solon},
  booktitle={Proceedings of the Conference on Fairness, Accountability, and Transparency},
  pages={39--48},
  year={2019}
}

@article{chouldechova2018frontiers,
  title={The frontiers of fairness in machine learning},
  author={Chouldechova, Alexandra and Roth, Aaron},
  journal={arXiv preprint arXiv:1810.08810},
  year={2018}
}

@incollection{narayanan2026limits,
  title={What if Algorithmic Fairness is a Category Error?},
  author={Narayanan, Arvind},
  booktitle={Contemporary Debates in the Ethics of Artificial Intelligence},
  editor={Nyholm, Sven and Kasirzadeh, Atoosa and Zerilli, John},
  pages={77--96},
  year={2026},
  publisher={Wiley-Blackwell},
  isbn={9781394258819}
}

@article{wang2024against,
  title={Against predictive optimization: On the legitimacy of decision-making algorithms that optimize predictive accuracy},
  author={Wang, Angelina and Kapoor, Sayash and Barocas, Solon and Narayanan, Arvind},
  journal={ACM Journal on Responsible Computing},
  volume={1},
  number={1},
  pages={1--45},
  year={2024},
  publisher={ACM New York, NY}
}

@article{johndrow2019algorithm,
  title={An algorithm for removing sensitive information},
  author={Johndrow, James E and Lum, Kristian},
  journal={The Annals of Applied Statistics},
  volume={13},
  number={1},
  pages={189--220},
  year={2019},
  publisher={JSTOR}
}

@inproceedings{guerdan2023ground,
  title={Ground (less) truth: A causal framework for proxy labels in human-algorithm decision-making},
  author={Guerdan, Luke and Coston, Amanda and Wu, Zhiwei Steven and Holstein, Kenneth},
  booktitle={Proceedings of the 2023 ACM Conference on Fairness, Accountability, and Transparency},
  pages={688--704},
  year={2023}
}

@inproceedings{jacobs2021measurement,
  title={Measurement and fairness},
  author={Jacobs, Abigail Z and Wallach, Hanna},
  booktitle={Proceedings of the 2021 ACM conference on fairness, accountability, and transparency},
  pages={375--385},
  year={2021}
}

@inproceedings{fogliato2021validity,
  title={On the validity of arrest as a proxy for offense: Race and the likelihood of arrest for violent crimes},
  author={Fogliato, Riccardo and Xiang, Alice and Lipton, Zachary and Nagin, Daniel and Chouldechova, Alexandra},
  booktitle={Proceedings of the 2021 AAAI/ACM Conference on AI, Ethics, and Society},
  pages={100--111},
  year={2021}
}

@book{alexander2011new,
  author    = {Alexander, Michelle},
  title     = {The New Jim Crow: Mass Incarceration in the Age of Colorblindness},
  year      = {2011},
  publisher = {The New Press},
  address   = {New York}
}

@article{Holm1979,
  author    = {Holm, Sture},
  title     = {A Simple Sequentially Rejective Multiple Test Procedure},
  journal   = {Scandinavian Journal of Statistics},
  volume    = {6},
  number    = {2},
  pages     = {65--70},
  year      = {1979},
  publisher = {Wiley}
}

@article{DresselFarid2018,
  author  = {Dressel, Julia and Farid, Hany},
  title   = {The Accuracy, Fairness, and Limits of Predicting Recidivism},
  journal = {Science Advances},
  volume  = {4},
  number  = {1},
  pages   = {eaao5580},
  year    = {2018}
}

@article{Flores2016,
  author  = {Flores, Anthony W. and Bechtel, Kristin and Lowenkamp, Christopher T.},
  title   = {False Positives, False Negatives, and False Analyses: A Rejoinder to {COMPAS} Critique},
  journal = {Federal Probation},
  volume  = {80},
  number  = {2},
  pages   = {38--46},
  year    = {2016}
}

@article{autor2003rise,
  author  = {Autor, David H.},
  title   = {Outsourcing at Will: The Contribution of Unjust Dismissal Doctrine to the Growth of Employment Outsourcing},
  journal = {Journal of Labor Economics},
  volume  = {21},
  number  = {1},
  pages   = {1--42},
  year    = {2003},
  doi     = {10.1086/344122}
}

@article{card1994minimum,
  author  = {Card, David and Krueger, Alan B.},
  title   = {Minimum Wages and Employment: A Case Study of the Fast-Food Industry in {New Jersey} and {Pennsylvania}},
  journal = {American Economic Review},
  volume  = {84},
  number  = {4},
  pages   = {772--793},
  year    = {1994}
}

@article{hewitt1991beck,
  title={The {Beck Anxiety Inventory}: Psychometric properties in a community population},
  author={Hewitt, Paul L. and Norton, G. Ron},
  journal={Journal of Psychopathology and Behavioral Assessment},
  volume={13},
  number={4},
  pages={345--352},
  year={1991},
  doi={10.1007/BF00960446}
}

@article{fawkes2024fragility,
  title={The fragility of fairness: Causal sensitivity analysis for fair machine learning},
  author={Fawkes, Jake and Fishman, Nic and Andrews, Mel and Lipton, Zachary},
  journal={Advances in Neural Information Processing Systems},
  volume={37},
  pages={137105--137134},
  year={2024}
}

@inproceedings{mhasawade2024causal,
  title={A causal perspective on label bias},
  author={Mhasawade, Vishwali and D'Amour, Alexander and Pfohl, Stephen R},
  booktitle={Proceedings of the 2024 ACM Conference on Fairness, Accountability, and Transparency},
  pages={1282--1294},
  year={2024}
}

@article{li2024depression,
  title={Depression in Adults with Chronic Pain: A Systematic Review and Meta-Analysis},
  author={Li, Jingyi and Zhou, Yiqing and Ma, Jing and others},
  journal={JAMA Network Open},
  volume={7},
  number={12},
  pages={e2451856},
  year={2024},
  doi={10.1001/jamanetworkopen.2024.51856}
}

@inproceedings{chouldechova2018case,
  title={A case study of algorithm-assisted decision making in child maltreatment hotline screening decisions},
  author={Chouldechova, Alexandra and Benavides-Prado, Diana and Fialko, Oleksandr and Vaithianathan, Rhema},
  booktitle={Conference on fairness, accountability and transparency},
  pages={134--148},
  year={2018},
  organization={PMLR}
}

@article{chouldechova2017fair,
  title={Fair prediction with disparate impact: A study of bias in recidivism prediction instruments},
  author={Chouldechova, Alexandra},
  journal={Big data},
  volume={5},
  number={2},
  pages={153--163},
  year={2017},
  publisher={Mary Ann Liebert, Inc. 140 Huguenot Street, 3rd Floor New Rochelle, NY 10801 USA}
}

@inproceedings{kleinberg2017inherent,
  title={Inherent Trade-Offs in the Fair Determination of Risk Scores},
  author={Kleinberg, Jon and Mullainathan, Sendhil and Raghavan, Manish},
  booktitle={8th Innovations in Theoretical Computer Science Conference (ITCS 2017)},
  pages={43--1},
  year={2017},
  organization={Schloss Dagstuhl--Leibniz-Zentrum f{\"u}r Informatik}
}

@article{salaudeen2025measurement,
  title={Measurement to Meaning: A Validity-Centered Framework for AI Evaluation},
  author={Salaudeen, Olawale and Reuel, Anka and Ahmed, Ahmed and Bedi, Suhana and Robertson, Zachary and Sundar, Sudharsan and Domingue, Ben and Wang, Angelina and Koyejo, Sanmi},
  journal={arXiv preprint arXiv:2505.10573},
  year={2025}
}

@inproceedings{watson2023multi,
  title={Multi-target multiplicity: Flexibility and fairness in target specification under resource constraints},
  author={Watson-Daniels, Jamelle and Barocas, Solon and Hofman, Jake M and Chouldechova, Alexandra},
  booktitle={Proceedings of the 2023 ACM Conference on Fairness, Accountability, and Transparency},
  pages={297--311},
  year={2023}
}

@inproceedings{coston2020counterfactual,
  title={Counterfactual risk assessments, evaluation, and fairness},
  author={Coston, Amanda and Mishler, Alan and Kennedy, Edward H and Chouldechova, Alexandra},
  booktitle={Proceedings of the 2020 conference on fairness, accountability, and transparency},
  pages={582--593},
  year={2020}
}

@article{rambachan2022robust,
  title={Robust design and evaluation of predictive algorithms under unobserved confounding},
  author={Rambachan, Ashesh and Coston, Amanda and Kennedy, Edward},
  journal={arXiv preprint arXiv:2212.09844},
  year={2022}
}

@article{kleinberg2018human,
  title={Human decisions and machine predictions},
  author={Kleinberg, Jon and Lakkaraju, Himabindu and Leskovec, Jure and Ludwig, Jens and Mullainathan, Sendhil},
  journal={The quarterly journal of economics},
  volume={133},
  number={1},
  pages={237--293},
  year={2018},
  publisher={Oxford University Press}
}

@inproceedings{edwards2016censoring,
  title={Censoring representations with an adversary},
  author={Edwards, Harrison and Storkey, Amos},
  booktitle={International Conference on Learning Representations (ICLR)},
  year={2016}
}

@inproceedings{madras2018learning,
  title={Learning adversarially fair and transferable representations},
  author={Madras, David and Creager, Elliot and Pitassi, Toniann and Zemel, Richard},
  booktitle={International Conference on Machine Learning (ICML)},
  pages={3384--3393},
  year={2018}
}

@inproceedings{zhang2018mitigating,
  title={Mitigating unwanted biases with adversarial learning},
  author={Zhang, Brian Hu and Lemoine, Blake and Mitchell, Margaret},
  booktitle={AAAI/ACM Conference on AI, Ethics, and Society},
  pages={335--340},
  year={2018}
}

@inproceedings{kawakami2024situate,
  title={The Situate AI Guidebook: Co-Designing a Toolkit to Support Multi-Stakeholder Early-stage Deliberations Around Public Sector AI Proposals},
  author={Kawakami, Anna and Coston, Amanda and Zhu, Haiyi and Heidari, Hoda and Holstein, Kenneth},
  booktitle={Proceedings of the CHI Conference on Human Factors in Computing Systems},
  year={2024}
}

@book{brown2015cfa,
  title={Confirmatory Factor Analysis for Applied Research},
  author={Brown, Timothy A.},
  edition={2},
  year={2015},
  publisher={Guilford Press}
}

@article{darlington1971another,
  title={Another look at "cultural fairness"},
  author={Darlington, Richard B},
  journal={Journal of Educational Measurement},
  volume={8},
  number={2},
  pages={71--82},
  year={1971}
}

@book{bollen1989sem,
  title={Structural Equations with Latent Variables},
  author={Bollen, Kenneth A.},
  year={1989},
  publisher={Wiley}
}

@article{westen2003quantifying,
  title={Quantifying construct validity: Two simple measures},
  author={Westen, Drew and Rosenthal, Robert},
  journal={Journal of Personality and Social Psychology},
  volume={84},
  number={3},
  pages={608--618},
  year={2003}
}

@book{cohen2003applied,
  title     = {Applied Multiple Regression/Correlation Analysis for the Behavioral Sciences},
  author    = {Cohen, Jacob and Cohen, Patricia and West, Stephen G. and Aiken, Leona S.},
  edition   = {3},
  year      = {2003},
  publisher = {Lawrence Erlbaum Associates},
  address   = {Mahwah, NJ}
}

@article{tversky1971belief,
  title={Belief in the law of small numbers},
  author={Tversky, Amos and Kahneman, Daniel},
  journal={Psychological Bulletin},
  volume={76},
  number={2},
  pages={105--110},
  year={1971},
  publisher={American Psychological Association}
}

@article{griffin1992weighing,
  title={The weighing of evidence and the determinants of confidence},
  author={Griffin, Dale and Tversky, Amos},
  journal={Cognitive Psychology},
  volume={24},
  number={3},
  pages={411--435},
  year={1992},
  publisher={Elsevier}
}

@article{ramdas2023game,
  title={Game-Theoretic Statistics and Safe Anytime-Valid Inference},
  author={Ramdas, Aaditya and Gr{\"u}nwald, Peter and Vovk, Vladimir and Shafer, Glenn},
  journal={Statistical Science},
  volume={38},
  number={4},
  pages={576--601},
  year={2023},
  publisher={Institute of Mathematical Statistics}
}

@article{waudbysmith2024estimating,
  title={Estimating Means of Bounded Random Variables by Betting},
  author={Waudby-Smith, Ian and Ramdas, Aaditya},
  journal={Journal of the Royal Statistical Society Series B: Statistical Methodology},
  volume={86},
  number={1},
  pages={1--27},
  year={2024},
  publisher={Oxford University Press}
}

@article{fithian2022conditional,
  title={Conditional Calibration for False Discovery Rate Control under Dependence},
  author={Fithian, William and Lei, Lihua},
  journal={Annals of Statistics},
  volume={50},
  number={6},
  pages={3091--3118},
  year={2022},
  publisher={Institute of Mathematical Statistics}
}


\clearpage
\appendix
 \section{Parametric Alternative for Algorithm 1}
\label{app:parametric_alternative}

In the main text, we present a nonparametric version of our falsification procedure (Alg.~\ref{alg:single_proxy}) that uses the Wilcoxon signed-rank test.
This choice makes minimal assumptions about the distribution of loss differences, providing robustness across a wide range of settings.
However, when the loss differences $\Delta_i$ satisfy certain distributional assumptions, a parametric alternative---the paired t-test---can achieve higher statistical power.

\paragraph{When to use the t-test.} The paired t-test assumes that the loss differences $\Delta_i$ are approximately normally distributed.
In practice, this assumption should be assessed before choosing between the Wilcoxon test and the t-test.
We recommend the following diagnostic checks:

\begin{enumerate}
    \item \textbf{Visual inspection:} Examine a histogram or Q-Q plot of the loss differences $\Delta_i$.
The distribution should appear approximately symmetric and bell-shaped.
    \item \textbf{Formal normality tests:} Apply the Shapiro-Wilk test \mbox{
\citep{shapiro1965analysis} }\hskip0pt
or other goodness-of-fit tests \mbox{
\citep{d2017goodness} }\hskip0pt
to assess normality.
Note that with large samples, these tests may reject normality even for minor deviations that do not substantially affect t-test validity.
    \item \textbf{Outlier detection:} Check for extreme values that could unduly influence the t-test.
Common approaches include examining values beyond 1.5 or 3 interquartile ranges from the quartiles \mbox{
\citep{barnett1994outliers}}\hskip0pt
.
\end{enumerate}

When the loss differences appear approximately normal and free of influential outliers, the t-test is preferable due to its higher power.
When in doubt, we recommend the Wilcoxon signed-rank test since it remains valid under weaker assumptions.
For additional guidance on assessing normality and handling outliers, we refer the reader to \mbox{
\citet{Freedman2007}}\hskip0pt
.

\paragraph{Algorithm with normality check.} Algorithm~\ref{alg:single_proxy_parametric} presents the full version of the falsification procedure that includes both the parametric (t-test) and nonparametric (Wilcoxon) options.

\begin{algorithm}[h]
\caption{Falsification Procedure with Single Permissible Proxy (with Parametric Option)}
\label{alg:single_proxy_parametric}
\begin{algorithmic}[1]
\REQUIRE Evaluation dataset $\mathcal{D} = \{(\hat f(x_i), y_{i,1}, \tilde{y}_i)\}_{i=1}^n$, significance level $\alpha$
\ENSURE $\texttt{DISCRIMINANT}$ or $\texttt{INDISCRIMINATE (inconclusive)}$ 
\FOR{$y \in \{\tilde{Y}, Y_1\}$}
    \STATE Apply Platt scaling for $y$: Fit logistic regression $P(y=1 \mid \hat f(x)) = \frac{1}{1 + \exp(A_y \cdot \hat f(x) + B_y)}$ on a held-out calibration set using labels for $y$, then transform predictions to calibrated probabilities $\hat p_i^{y}$
\ENDFOR
\STATE Compute losses: $L_{i,\tilde{Y}} = \ell(\tilde{y}_i, \hat p_i^{\tilde{Y}})$ and $L_{i,Y_1} = \ell(y_{i,1}, \hat p_i^{Y_1})$ for $i = 1, \ldots, n$
\STATE Compute loss differences: $\Delta_i = L_{i,\tilde{Y}} - L_{i,Y_1}$ for $i = 1, \ldots, n$
\STATE Check normality and outliers in $\Delta_i$ 
\IF{data appears normal and without outliers}
    \STATE \textbf{Perform t-test on differences:} Test $H_0: \mathbb{E}[\Delta] \leq 0$ vs $H_1: \mathbb{E}[\Delta] > 0$
\ELSE
    \STATE \textbf{Perform Wilcoxon signed-rank test:} Test $H_0: \theta \leq 0$ vs $H_1: \theta > 0$ where $\theta$ is the location parameter of the distribution of the loss differences.
Under symmetry, the location parameter is the median loss difference.
    \STATE \hspace{1.5em}-- Compute $W = \sum_{i=1}^n \text{sign}(\Delta_i) \cdot rank(|\Delta_i|)$ 
    \STATE \hspace{1.5em}-- Compute p-value using $W$ and sample size $n$. 
\ENDIF
\IF{p-value $\leq \alpha$}
    \RETURN $\texttt{DISCRIMINANT}$
\ELSE
    \RETURN $\texttt{INDISCRIMINATE (inconclusive)}$
\ENDIF
\end{algorithmic}
\end{algorithm}

\section{Correlation analysis and visualizations for COMPAS experiments}
\label{app:compas_correlation_viz}

\subsection{Correlation Analysis}
\label{app:correlation_compas}
\begin{table}[h]
    \centering
    \caption{Correlation between COMPAS decile risk scores and each proxy on the COMPAS evaluation set.
    Colors indicate impermissible proxies (\textcolor{red!50!orange}{age $< 25$} and \textcolor{red!80!black}{race}).}
    \label{tab:correlation_compas}
    \small
    \begin{tabular}{lccc}
    \toprule
    \textbf{Measure} & \textbf{Re-arrest} & \textcolor{red!50!orange}{\textbf{Age $< 25$}} & \textcolor{red!80!black}{\textbf{Race (Black)}} \\
    \midrule
    Pearson & 0.3590 & \textcolor{red!50!orange}{0.2447} & \textcolor{red!80!black}{0.3060} \\
    Spearman's $\rho$ & 0.3533 & \textcolor{red!50!orange}{0.2619} & \textcolor{red!80!black}{0.3121} \\
    Kendall's $\tau$ & 0.3044 & \textcolor{red!50!orange}{0.2256} & \textcolor{red!80!black}{0.2689} \\
    \bottomrule
    \end{tabular}
    \end{table}

Table~\ref{tab:correlation_compas} reports Pearson, Spearman, and Kendall correlations between the COMPAS decile risk scores and each proxy on the evaluation set.
Correlation with the permissible proxy (re-arrest) is higher than correlations with the impermissible proxies, but across the board these correlations are quite low and it's unclear whether differences are statistically significant.
Our hypothesis-based method clarifies this by testing differences in predictive loss, finding that we fail to establish the discriminant validity COMPAS scores with respect to either race or age.

 \section{Ablation Studies}
\label{app:ablation}

We conduct ablation studies to understand the contribution of different components of our falsification procedure.

\subsection{Effect of Platt Scaling}

We examine whether Platt scaling calibration affects the falsification test results by comparing outcomes with and without this calibration step.
Table~\ref{tab:ablation_platt} summarizes the results.

\begin{table}[h]
\centering
\caption{Ablation study: Effect of Platt scaling on falsification test results.
Logistic regression results use Algorithm~1 (single permissible proxy); LAFTR results use Algorithm~2 (multiple permissible proxies).}
\label{tab:ablation_platt}
\small
\begin{tabular}{llllcc}
\toprule
\textbf{Model} & \textbf{Imperm.} & \textbf{Calibration} & \textbf{Mean Diff.} & \textbf{P-value} & \textbf{Result} \\
\midrule
\multicolumn{6}{l}{\emph{LSAC}} \\
Logistic reg. & Race & With Platt & $+0.50$ & $\approx 1$ & INDISCRIMINANT \\
Logistic reg. & Race & Without Platt & $-0.03$ & $\approx 0$ & DISCRIMINANT \\
Logistic reg. & Gender & With Platt & $-0.03$ & $\approx 0$ & DISCRIMINANT \\
Logistic reg. & Gender & Without Platt & $-0.07$ & $\approx 0$ & DISCRIMINANT \\
\addlinespace
LAFTR (adv=race) & Race & With Platt & --- & $\approx 1$ & INDISCRIMINANT \\
LAFTR (adv=race) & Race & Without Platt & $+0.04$ & $\approx 0$ & DISCRIMINANT \\
LAFTR (adv=race) & Gender & With Platt & --- & $\approx 0$ & DISCRIMINANT \\
LAFTR (adv=race) & Gender & Without Platt & $+0.02$ & $\approx 0$ & DISCRIMINANT \\
\addlinespace
LAFTR (adv=gender) & Gender & With Platt & --- & $\approx 0$ & DISCRIMINANT \\
LAFTR (adv=gender) & Gender & Without Platt & $+0.06$ & $\approx 0$ & DISCRIMINANT \\
LAFTR (adv=gender) & Race & With Platt & --- & $\approx 1$ & INDISCRIMINANT \\
LAFTR (adv=gender) & Race & Without Platt & $+0.03$ & $0.13$ & INDISCRIMINANT \\
\midrule
\multicolumn{6}{l}{\emph{COMPAS}} \\
COMPAS scores & Age $<$ 25 & With Platt & $+0.14$ & $\approx 1$ & INDISCRIMINANT \\
COMPAS scores & Age $<$ 25 & Without Platt & $+0.59$ & $\approx 1$ & INDISCRIMINANT \\
COMPAS scores & Race & With Platt & $-0.03$ & $\approx 0$ & DISCRIMINANT \\
COMPAS scores & Race & Without Platt & $-0.18$ & $\approx 0$ & DISCRIMINANT \\
\bottomrule
\end{tabular}
\end{table}

For COMPAS, the qualitative result (INDISCRIMINANT) is unchanged by Platt scaling.
For the logistic regression on LSAC, the qualitative result for race as the impermissible proxy \emph{changes} depending on whether Platt scaling is applied: without calibration the test returns DISCRIMINANT, but with proper calibration it returns INDISCRIMINANT.
The same pattern appears for the LAFTR model with the race adversary: without calibration, the test incorrectly returns DISCRIMINANT for race, but with Platt scaling it correctly returns INDISCRIMINANT.
This artifact arises from comparing losses computed from uncalibrated predictions against labels with very different base rates (race: 94\% white vs.\ GPA: $\sim$50\%).
Platt scaling calibrates predictions to each outcome separately, putting them on the same probability scale and enabling meaningful comparison.
These results demonstrate that proper calibration is essential for valid falsification testing---without it, models that are actually more aligned with race than with permissible outcomes would incorrectly appear to pass the test.
Notably, the LAFTR gender-adversary model tested against race returns INDISCRIMINANT both with and without Platt scaling, though the p-value is much less extreme without calibration ($p = 0.13$ vs.\ $p \approx 1$).

\subsection{Effect of Loss Function}

We compare results using log loss (binary cross-entropy) versus Brier score (squared error).
Table~\ref{tab:ablation_loss} summarizes the results.

\begin{table}[h]
\centering
\caption{Ablation study: Effect of loss function on falsification test results (Algorithm 1).}
\label{tab:ablation_loss}
\small
\begin{tabular}{lllccc}
\toprule
\textbf{Dataset} & \textbf{Impermissible} & \textbf{Loss Function} & \textbf{Mean Diff.} & \textbf{P-value} & \textbf{Result} \\
\midrule
LSAC & Race & Log loss & $+0.50$ & $\approx 1$ & INDISCRIMINANT \\
LSAC & Race & Brier score & $+0.19$ & $\approx 1$ & INDISCRIMINANT \\
LSAC & Gender & Log loss & $-0.03$ & $\approx 0$ & DISCRIMINANT \\
LSAC & Gender & Brier score & $-0.01$ & $\approx 0$ & DISCRIMINANT \\
\midrule
COMPAS & Age $<$ 25 & Log loss & $+0.14$ & $\approx 1$ & INDISCRIMINANT \\
COMPAS & Age $<$ 25 & Brier score & $+0.06$ & $\approx 1$ & INDISCRIMINANT \\
COMPAS & Race & Log loss & $-0.03$ & $\approx 0$ & DISCRIMINANT \\
COMPAS & Race & Brier score & $-0.01$ & $\approx 0$ & DISCRIMINANT \\
\bottomrule
\end{tabular}
\end{table}

The choice of loss function does not affect the qualitative conclusions for either dataset.
Log loss tends to produce larger magnitude differences because it penalizes confident incorrect predictions more severely.
The consistency of results across loss functions suggests our findings are robust to this methodological choice.

\section{Additional LSAC Results}
\label{app:lsac_additional_results}

\subsection{Negative Control}
\label{app:negative_control}

As a validation of our procedure, we apply the falsification test where we treat a permissible proxy as if it were impermissible.
If the procedure is working correctly, we should fail to establish that the model predicts other permissible proxies better than the one we designate as ``impermissible.'' We conduct this negative control in the law school admissions setting, using the LSAC data.

We apply Alg.1 with cumulative GPA (in reality a permissible proxy) as the impermissible proxy and first-year GPA as the permissible proxy.
As expected, the test returns \texttt{INDISCRIMINANT} with $p \approx 1$, indicating we cannot establish that the model predicts first-year GPA better than cumulative GPA. 

\subsection{Alternative Outcome Specification}
\label{app:bar_passage_model}

To assess whether our findings are sensitive to the choice of modeled outcome, we train an alternative model using bar passage as the outcome instead of first-year GPA.
Table~\ref{tab:performance_metrics_bar} shows the performance metrics for this model.

\begin{table}[h]
\centering
\caption{Performance metrics for a model trained to predict bar passage (instead of first-year GPA).
The pattern is similar: race has the highest AUC (0.90), while gender is near-random (0.50).}
\label{tab:performance_metrics_bar}
\small
\begin{tabular}{l@{\hspace{0.5em}}ccccccc}
\toprule
\textbf{Proxy} & \textbf{AUC} & \textbf{AU PR} & \textbf{MSE} & \textbf{PPV Top 2\%} & \textbf{PPV Top 10\%} & \textbf{PPV Top 50\%} & \textbf{PPV Top 75\%} \\
\midrule
1styearGPA & 0.6257 & 0.5993 & 0.3991 & 0.6596 & 0.6368 & 0.5905 & 0.5526 \\
GPA & 0.6538 & 0.6225 & 0.3959 & 0.6809 & 0.6581 & 0.6132 & 0.5658 \\
Pass Bar & 0.7626 & 0.9639 & 0.0771 & 0.9894 & 0.9915 & 0.9632 & 0.9472 \\
\textcolor{red!80!black}{Race} & \textcolor{red!80!black}{0.8953} & \textcolor{red!80!black}{0.9908} & \textcolor{red!80!black}{0.0447} & \textcolor{red!80!black}{1.0000} & \textcolor{red!80!black}{1.0000} & \textcolor{red!80!black}{0.9923} & \textcolor{red!80!black}{0.9852} \\
\textcolor{red!50!orange}{Gender} & \textcolor{red!50!orange}{0.4969} & \textcolor{red!50!orange}{0.5764} & \textcolor{red!50!orange}{0.3638} & \textcolor{red!50!orange}{0.6170} & \textcolor{red!50!orange}{0.5983} & \textcolor{red!50!orange}{0.5665} & \textcolor{red!50!orange}{0.5686} \\
\bottomrule
\end{tabular}
\end{table}

The results mirror those of the first-year GPA model: race has the highest AUC (0.90), while gender remains near-random (0.50). 

\subsection{Additional Impermissible Proxies}
\label{app:lsac_additional}

In addition to the race-based and gender-based falsification analysis presented in the main text, we apply our falsification procedure to the same LSAC predictive model using family income as an alternative impermissible proxy.
\subsubsection{Family Income as Impermissible Proxy}

We similarly apply our falsification procedure using family income as the impermissible proxy.
Family income in the LSAC dataset is encoded as an ordinal variable with categories 1--5.
We binarize this variable by defining high income as the highest category (category 5), which comprises approximately 8\% of applicants.
The test returns \texttt{INDISCRIMINANT} with p-value $\approx 1$, indicating we cannot establish that the model predicts the permissible proxies better than it predicts family income.
Although the raw AUC for family income (0.54) is lower than for GPA (0.65), after Platt scaling calibration, the model predicts family income with lower loss than GPA for 92\% of observations.
This result fails to establish discriminant validity with respect to family income.

\subsection{LAFTR Models for LSAC data: Architectural details and baseline performance metrics}
\label{app:laftr_metrics}

\paragraph{Model architecture} LAFTR learns a fair representation by training an encoder--classifier architecture adversarially against a discriminator that attempts to predict the sensitive attribute from the learned representation.
We train two variants: one with a demographic-parity adversary targeting race and one targeting gender.
Both models use a three-layer encoder consisting of a hidden layer with 64 nodes and an output layer with 32 nodes and ReLU activation which we denote with the shorthand (input $\to$ 64 $\to$ 32 ReLU); a classifier (32 $\to$ 64 $\to$ 1), an adversary (32 $\to$ 64 $\to$ 1), and a reconstructor (32 $\to$ 64 $\to$ input), trained for 150 epochs with Adam (lr = $10^{-3}$), adversary weight 1.0, and reconstruction weight 0.1.

\paragraph{Baseline metrics} Table~\ref{tab:laftr_metrics} reports the same performance metrics as Table~\ref{tab:performance_metrics} for the two LAFTR models \citep{madras2018learning}.
If we wanted to make a determination about discriminant validity based on the information in this table, our conclusion may differ depending on which metric we choose.
MSE suggests that the model performs better on permissible proxies than on race (although we might debate whether this difference is statistically significant), whereas the other metrics suggest that the model performs as good or even better on race than on the permissible proxies.
This illustrates the limitations of trying to assess discriminant validity by descriptively comparing differences in performance metrics -- our conclusions are brittle, depending on metric choice and how we determine what a significant disparity in metric is.

\begin{table}[h]
\centering
\caption{Performance metrics on the LSAC dataset for LAFTR models \citep{madras2018learning}.
Colors indicate impermissible proxies (\textcolor{red!80!black}{race} and \textcolor{red!50!orange}{gender}).}
\label{tab:laftr_metrics}
\small
\begin{tabular}{l@{\hspace{0.5em}}ccccccc}
\toprule
\textbf{Proxy} & \textbf{AUC} & \textbf{AU PR} & \textbf{MSE} & \textbf{PPV Top 2\%} & \textbf{PPV Top 10\%} & \textbf{PPV Top 50\%} & \textbf{PPV Top 75\%} \\
\midrule
\multicolumn{8}{l}{\emph{LAFTR (adversary = race)}} \\
1styearGPA & 0.5961 & 0.5774 & 0.2439 & 0.5806 & 0.6317 & 0.5749 & 0.5365 \\
GPA & 0.6244 & 0.5939 & 0.2398 & 0.5914 & 0.6231 & 0.5967 & 0.5499 \\
Pass Bar & 0.6461 & 0.9305 & 0.2682 & 0.8387 & 0.9358 & 0.9431 & 0.9284 \\
\textcolor{red!80!black}{Race} & \textcolor{red!80!black}{0.6377} & \textcolor{red!80!black}{0.9495} & \textcolor{red!80!black}{0.2725} & \textcolor{red!80!black}{0.8495} & \textcolor{red!80!black}{0.9400} & \textcolor{red!80!black}{0.9623} & \textcolor{red!80!black}{0.9541} \\
\textcolor{red!50!orange}{Gender} & \textcolor{red!50!orange}{0.4911} & \textcolor{red!50!orange}{0.5698} & \textcolor{red!50!orange}{0.2628} & \textcolor{red!50!orange}{0.5806} & \textcolor{red!50!orange}{0.5953} & \textcolor{red!50!orange}{0.5646} & \textcolor{red!50!orange}{0.5616} \\
\midrule
\multicolumn{8}{l}{\emph{LAFTR (adversary = gender)}} \\
1styearGPA & 0.6252 & 0.6018 & 0.2380 & 0.6667 & 0.6510 & 0.5912 & 0.5519 \\
GPA & 0.6558 & 0.6307 & 0.2317 & 0.7527 & 0.6916 & 0.6134 & 0.5656 \\
Pass Bar & 0.7627 & 0.9636 & 0.2571 & 1.0000 & 0.9829 & 0.9649 & 0.9463 \\
\textcolor{red!80!black}{Race} & \textcolor{red!80!black}{0.8934} & \textcolor{red!80!black}{0.9904} & \textcolor{red!80!black}{0.2549} & \textcolor{red!80!black}{1.0000} & \textcolor{red!80!black}{0.9979} & \textcolor{red!80!black}{0.9927} & \textcolor{red!80!black}{0.9854} \\
\textcolor{red!50!orange}{Gender} & \textcolor{red!50!orange}{0.4956} & \textcolor{red!50!orange}{0.5599} & \textcolor{red!50!orange}{0.2689} & \textcolor{red!50!orange}{0.4301} & \textcolor{red!50!orange}{0.5396} & \textcolor{red!50!orange}{0.5634} & \textcolor{red!50!orange}{0.5679} \\
\bottomrule
\end{tabular}
\end{table}

\subsection{Nonparametric Correlation Measures}
\label{app:correlation_nonparametric}

Table~\ref{tab:correlation_table_nonparametric} reports Spearman's $\rho$ and Kendall's $\tau$ between model predictions and each proxy on the LSAC evaluation set, complementing the Pearson correlations in Table~\ref{tab:correlation_table}.
The substantive conclusions are identical across all three measures of correlation (Pearson's, Spearman's, and Kendall's).

\begin{table}[h]
\centering
\caption{Spearman and Kendall correlations between model predictions and each proxy on the LSAC evaluation set.
All models are trained on LSAT and undergraduate GPA to predict first-year GPA.
Results are substantively identical to the Pearson correlations in Table~\ref{tab:correlation_table}.}
\label{tab:correlation_table_nonparametric}
\small
\begin{tabular}{lccccc}
\toprule
\textbf{Model} & \textbf{1styearGPA} & \textbf{GPA} & \textbf{Pass Bar} & \textcolor{red!80!black}{\textbf{Race}} & \textcolor{red!50!orange}{\textbf{Gender}} \\
\midrule
\multicolumn{6}{l}{\emph{Spearman's $\rho$}} \\
Logistic regression & 0.2175 & 0.2680 & 0.2687 & \textcolor{red!80!black}{0.3321} & \textcolor{red!50!orange}{$-$0.0033} \\
LAFTR (adv=race) & 0.1665 & 0.2155 & 0.1493 & \textcolor{red!80!black}{0.1158} & \textcolor{red!50!orange}{$-$0.0153} \\
LAFTR (adv=gender) & 0.2168 & 0.2698 & 0.2685 & \textcolor{red!80!black}{0.3309} & \textcolor{red!50!orange}{$-$0.0075} \\
\midrule
\multicolumn{6}{l}{\emph{Kendall's $\tau$}} \\
Logistic regression & 0.1779 & 0.2192 & 0.2198 & \textcolor{red!80!black}{0.2716} & \textcolor{red!50!orange}{$-$0.0027} \\
LAFTR (adv=race) & 0.1362 & 0.1763 & 0.1221 & \textcolor{red!80!black}{0.0947} & \textcolor{red!50!orange}{$-$0.0125} \\
LAFTR (adv=gender) & 0.1774 & 0.2207 & 0.2196 & \textcolor{red!80!black}{0.2706} & \textcolor{red!50!orange}{$-$0.0061} \\
\bottomrule
\end{tabular}
\end{table}

\section{Additional empirical results for COMPAS}
\label{app:compas_additional_results}

\subsection{Falsification of a Trained Predictive Model}

In addition to evaluating the existing COMPAS risk scores, we also trained our own logistic regression model to predict two-year recidivism and applied our falsification procedure to this model.
This analysis allows us to assess whether our findings are specific to the proprietary COMPAS scores or generalize to other predictive models in this setting.

We train a logistic regression model to predict two-year recidivism (\texttt{two\_year\_recid}) using the following predictive features from the COMPAS dataset: sex, age, juvenile felony count (\texttt{juv\_fel\_count}), juvenile misdemeanor count (\texttt{juv\_misd\_count}), juvenile other count (\texttt{juv\_other\_count}), prior offenses count (\texttt{priors\_count}), and charge degree indicators (\texttt{c\_charge\_degree\_F} and \texttt{c\_charge\_degree\_M}).
We exclude race-related variables (including one-hot encoded race columns) and age category indicators (since age categories are used to define the impermissible proxy).
The dataset is split into training (60\%), calibration (20\%), and evaluation (20\%) sets.
The model is trained on the training set, Platt scaling calibration is performed on the calibration set, and all falsification tests are conducted on the evaluation set.

We apply Algorithm 1 (single permissible proxy test) to assess whether the model predicts the impermissible proxy (age < 25) better than it predicts the modeled outcome (two-year recidivism). 

Our method obtains a p-value of $1-1e-5 \approx 1$, indicating that we are not able to reject the null hypothesis that the model predicts the modeled outcome better than it predicts which defendants are less than 25.
This finding is consistent with our analysis of the existing COMPAS risk scores and suggests that the inability to establish discriminant validity with respect to age is not unique to the proprietary COMPAS scoring system, but rather reflects a broader challenge in recidivism prediction models.


\end{document}